\documentclass[preprint2]{aastex}
\usepackage{graphicx,epsfig,times}

\slugcomment{Accepted to the Astronomical Journal -- 30 April 2004}

\shorttitle{Quasars as Absorption Probes}
\shortauthors{McIntosh, Impey, \& Petry}

\begin{document}


\title{Quasars as Absorption Probes of the J0053+1234 Region\footnote{Observations reported here were obtained at
Kitt Peak National Observatory, National Optical Astronomy Observatories,
which is operated by the Association of Universities for Research in Astronomy,
Inc., under cooperative agreement with the National Science Foundation.}}



\author{Daniel H. McIntosh\footnote{Present address: University of Massachusetts, Lederle Graduate Research Tower, Amherst, MA 01003}, Chris D. Impey, and Catherine E. Petry} 
\affil{Steward Observatory, University of Arizona, 933 North Cherry Avenue, Tucson, AZ  85721}
\email{dmac@hamerkop.astro.umass.edu, cimpey@as.arizona.edu, cpetry@as.arizona.edu}

\begin{abstract}
We present $UBRI$ photometry and spectra for 60 quasars found within
one square degree centered on the J0053+1234 region, which has 
been the subject of the Caltech Faint Galaxy Redshift Survey.  
Candidate quasars 
were selected by their ultraviolet excess with respect to the stellar 
locus, and confirmed spectroscopically.  The quasars span a wide range 
in brightness ($17.5<B<21.6$) and redshift ($0.43<z<2.38$).  These new 
quasars comprise a grid of absorption probes that can be used to study 
large-scale structure as well as the correlation between luminous galaxies, 
non-luminous halos, and Lyman-$\alpha$ absorbers in the direction of the 
deep pencil-beam galaxy survey.  Spectra of 14 emission line galaxies
found using the same technique are also presented.

\end{abstract}



\section{Introduction}

Pencil-beam galaxy redshift surveys are enormously useful 
in defining the evolution and clustering of luminous matter out 
to redshifts $z \gtrsim 1$.  An important complement to such surveys is
a network of quasar absorbers within the same volume, which allows the detection
of matter in absorption that would be impossible to detect in emission.
Leveraging a large investment of resources from the Hubble Space Telescope
({\it HST}), the two Hubble Deep Fields 
\citep[HDF-N and HDF-S][]{williams96,williams00} have
attracted an international campaign of multi-wavelength observations.
The more recent
Ultra-Deep Field (UDF) is attracting similar attention. The Caltech Faint
Galaxy Redshift Survey (CFGRS) has devoted a concentrated effort
to two fields, using the Low Resolution Imaging Spectrograph at
the Keck Observatory \citep{cohen96}. One CFGRS field is the 
northern Hubble Deep Field, and the other is a previously unstudied 
field centered at $\alpha = 00^{\rm h} 53^{\rm m} 23^{\rm s}$, 
$\delta = 12\degr 33\arcmin 58\arcsec$ (J2000). Initial 
spectroscopy has been published by \citet{cohen99a,cohen99b}, 
and a clustering analysis based on both fields has also been 
presented \citep{hogg00}. These studies concentrate on the distribution
and evolution of luminous matter along this sightline, and will 
eventually reveal the evolutionary properties of galaxies of 
different Hubble types, delineate large-scale structure in redshift,
and help define the cosmic history of star formation.
Here we provide a grid of quasar probes, which will allow an examination
of the cold, diffuse, and dark components of the universe along the 
J0053+1234 pencil beam.

Regardless of the depth of an imaging survey, 
it can reveal only the luminous parts of galaxies, which contain about
10--15\% of the baryons in the local universe, which in turn form
only about 15\% of the matter content of the universe \citep{turner01}. 
Studying the cold, diffuse, and dark components of the 
universe in a cosmological volume centered on J0053+1234 provides an important
complement to the study of the luminous matter content. This can 
be accomplished by absorption line spectroscopy, using distant 
quasars as probes.  For example, galaxy halos can be detected indirectly using 
the \ion{C}{4} $\lambda\lambda$1548, 1550 and \ion{Mg}{2}
$\lambda\lambda$2796, 2800 doublets over the entire range $0 < z 
< 4$ \citep[e.g.][]{meylan95}.  Furthermore, Lyman-$\alpha$ 
absorbers are as numerous
as galaxies and they effectively trace the gravitational potential 
of the underlying dark matter \citep{miralda96,croft98}.
Given a sufficiently bright background quasar, 
absorbers can be detected with an efficiency that does not depend 
on redshift.  In this respect, quasar absorption probes have an advantage
over galaxy surveys that are always affected
by, and sometimes compromised by, effects such as 
cosmological $(1+z)^4$ dimming, $k$-corrections, and morphological
selection that depends on redshift.

The detection of a network of quasar absorbers in a volume that
encompasses a deep pencil-beam survey allows measurements of large-scale
structure that can be used to relate luminous baryons in galaxies 
to baryons
that are largely intergalactic. Individual quasar sightlines show that
metals such as
\ion{C}{4} and \ion{Mg}{2} have correlation power on scales up to 
100 $h^{-1}_{100}$ Mpc \citep{loh01}, and multiple
sightlines have been used to trace out three-dimensional structures
on even larger scales \citep{dinshaw96,williger96}.
Locally, a grid of quasar probes have been used to trace out diffuse
gas structure in three dimensions and relate it to the galaxy
distribution in the direction of the galactic poles \citep{vandenberk99}
and in the direction of the Virgo cluster \citep{impey99}.
There is evidence that Lyman-$\alpha$ absorbers are significantly 
but weakly clustered on scales of 20-30 $h^{-1}_{100}$ Mpc 
\citep{williger00,liske00}.  Deep pencil-beam galaxy redshift
surveys have shown that around half of the bright galaxies lie in high contrast
structures with line of sight separations of 50-300 $h^{-1}_{100}$ 
Mpc \citep{cohen96,cohen99a}. Thus, the spatial relationship 
between quasar absorbers and luminous galaxies can be studied
as a function of redshift. Gas dynamical simulations indicate 
that low column density absorbers more closely reflect the
underlying dark matter mass distribution than galaxies \citep{cen98}.

We have already published first results
on a quasar survey in an area encompassing the HDF-N
\citep{liu99,vandenberk00}, and we are completing
a multi-color search for quasars in a region centered on the HDF-S 
\citep[see also][]{palunas00}. In this paper, we present the 
observations that have led to the discovery of 60 quasars in a 
one square degree field centered on the J0053+1234 region of
\citet{cohen99a,cohen99b}. 
The galaxy redshift survey covered a region 14.6 arcmin$^{2}$ at the center
of the field surveyed for quasars.
In all three cases, quasar selection is a prelude to absorption line 
spectroscopy and a comparison of the spatial distribution of
absorbers and galaxies. Although quasars as faint as $B\approx21.5$
have traditionally been too challenging for absorption line work,
a new generation of large ground-based telescopes and multi-object
spectrographs will enable the follow-up observations required for
this sample.

We describe the imaging and multi-color photometry leading
to the selection of quasar candidates in \S2, and the quasar candidate
selection in \S3.  In \S4 we describe the spectroscopy. In \S5
we summarize the yield of confirmed quasars, the reliability of 
the redshifts, and present a smaller number of emission line galaxies
discovered in the same survey.
We close with brief comments on future work on this
region.
A second paper (Petry, Impey, \& McIntosh, in preparation)  will
present additional quasar candidates for the J0053+1234 field
derived from deeper $U$ imaging of the central half square degree,
as well as confirmed quasars in the entire field 
from two recent observing runs.
Additional papers will present photometry, quasar candidate
lists, and confirmed quasars for both the Hubble
Deep Fields, North and South.  Spectra have been obtained for quasars 
in all three fields that are of sufficient quality to measure strong
Lyman-$\alpha$ absorbers,
so these future papers will also include a first comparison between
quasar absorbers and galaxies in the three pencil-beam surveys.

\section{Photometry}

\subsection{Observations}

We used the Kitt Peak National Observatory (KPNO) 0.9-meter Telescope with the 
CCD Mosaic Wide-Field Imager \citep{boroson94,muller98} to acquire $UBRI$ 
observations of the one square-degree region centered on the J0053+1234 region 
($\rm \alpha = 00^{h}53^{m}23.2^{s}, \delta= +12\degr 33\arcmin 58\arcsec$)
during 1998 September 30 -- October 2 UT, under generally good observing 
conditions.

\begin{figure}[hp]
  \vskip 5truecm
  \centering
  \mbox{
\includegraphics[scale=.7, angle=270]{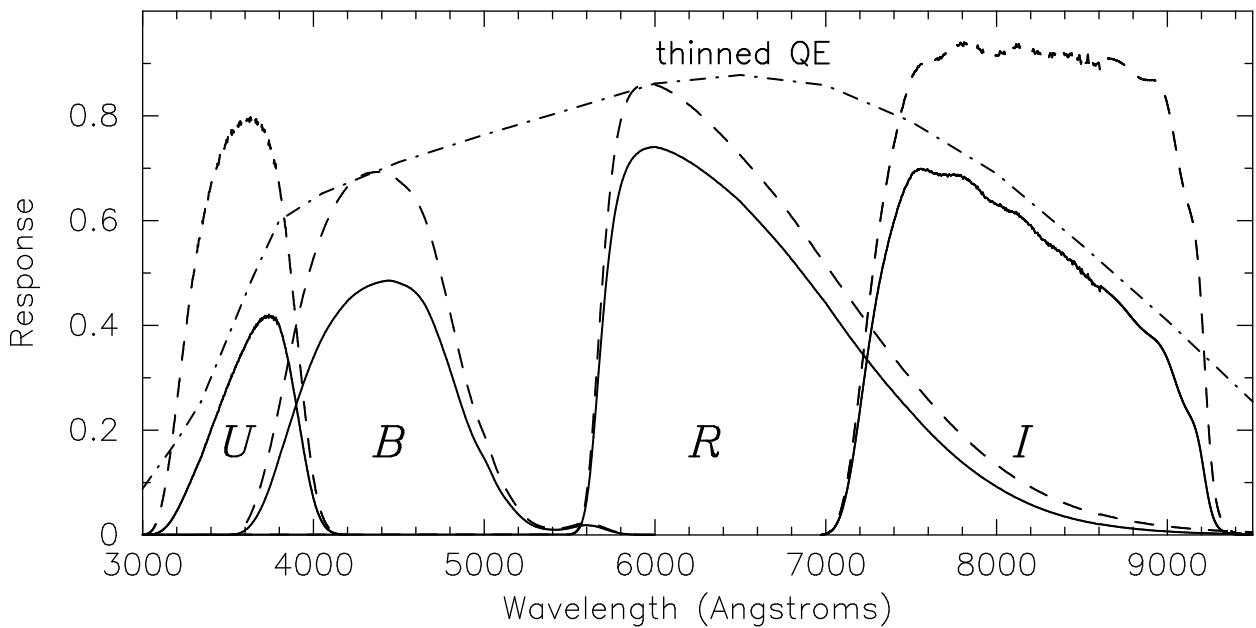}
  }
\caption
{Average KPNO Mosaic response in the $UBRI$ passbands (solid lines).
The dash-dotted
line represents the average quantum efficiency (QE)
of the eight Mosaic Loral CCD's.
The transmission of each filter is given by the dashed
lines.  The total response is the combination of the filter transmission
and QE.  The $U$-band response of the system used for our observations drops off
rapidly blueward of $\sim3750$~\AA.
}
\label{response}
\end{figure}

The Mosaic imager has eight $2048$\,$\times$\,$4096$ pixel anti-reflection
coated SITe 
Loral chips.  The combined field of view spans $59\arcmin \times 59\arcmin$,
and has a pixel scale of $0\farcs43$ (15 $\micron$ pixels).
The readout noise for this camera is $5.66 e^-$, the dark 
current is negligible ($\sim15 e^-$~hour$^{-1}$), and the average single
chip gain is $2.86 e^-$~ADU$^{-1}$.  Each CCD has been thinned for detecting 
$U$-band photons; however, the quantum efficiency (QE) still falls off rapidly 
blueward of 4000 \AA.  Thus, the spectral response in the $U$-band is not a 
perfect match to standard \citet{johnson66} $U$-band.  The Mosaic camera 
employs a large (5.75 inches square) par-focal filter set.  Each filter's 
transmission and response, calculated from the QE, is plotted in 
Figure \ref{response}.

An area of sky centered on the J0053+1234 region was imaged in each
passband using a standard dither pattern with five separate pointings
a few arcminutes apart to remove inter-chip gaps, CCD defects
and cosmic rays.  This dither pattern ensured at least $80\%$ of the 
maximum exposure for all regions of a final combined image of 5 exposures.
The total integration times were 420, 110, 30, and 70 minutes for the
$U$, $B$, $R$ and $I$ bands, respectively.  A variety of short and long 
exposure times were used for individual $U$ and $I$ frames to avoid
saturation of the brightest objects. 
Unfortunately, we did not experience perfectly photometric conditions during our
imaging campaign.  Nevertheless, for relative photometric calibration
we imaged \citet{landolt92} 
standard star cluster fields, over a range of airmasses, 
at least three times during each night.  The overall photometric zero point
consistency was $\sim 10$\%. 
Quasar selection is governed by relative colors with
respect to the stellar locus; therefore, zero point determination did not
limit our ability to identify quasar candidates with this photometry.
We summarize the observations in Table \ref{obslog}.

\subsection{Data Reduction}
To select quasar candidates in the J0053+1234 region for spectroscopic 
follow-up, we require good photometric
uniformity across our deep, wide-field images.  Homogeneous photometry also
requires procedures to deal with bad pixels, cosmic rays,
and gaps between CCD's in each passband.  Achieving such high quality images
requires stacking dithered Mosaic frames, which in turn 
places high demands on the initial data reduction steps
\footnote{NOAO CCD Mosaic Imager User 
Manual (hereafter MosManual), Version Sept. 15, 2000,
G. Jacoby; http://www.noao.edu/kpno/mosaic/manual/index.html.}.  
In particular, the 
data must be well-flattened and carefully corrected for photometric effects of
the variable pixel scale.  The data are reduced
with a customized reduction pipeline that uses the IRAF\footnote{IRAF is
distributed by the National Optical Astronomical Observatories, which are
operated by AURA, Inc. under contract to the NSF.}
environment and adheres to standard image reduction techniques.

We perform basic reduction of the individual Mosaic frames using the IRAF
{\it mscred} package.  This software allows image processing to be performed on
multi-chip exposures as if they were single CCD frames.  For each single
Mosaic exposure, we trim and debias the eight individual chip images
separately.  A small correction ($<0.3\%$)
is necessary due to cross talk between
pairs of adjacent chips sharing the same electronics in
the Mosaic detector.  This correction subtracts a predetermined fraction
of the adjacent chip's $(i,j)^{\rm th}$ pixel value from pixel $(i,j)$ of
the current chip.  We then remove an averaged zero frame from 
each Mosaic image.
The thinned Mosaic chips require no dark correction.  Roughly $0.4\%$ of the
full array of $8192\times 8192$ pixels are bad; these are flagged and included
in the mask frames during the final image combination.

An important step towards achieving precise photometry is the determination and
removal of the response function of the individual CCD's -- {\it i.e.}
flat fielding the data.
Traditionally, dividing each exposure by a uniformally
illuminated blank frame will produce an image that has a uniform and
flat appearance.  Yet,
the Mosaic imagers have pixel scales that decrease roughly quadratically
such that an individual pixel in a field corner is 6\% smaller,
and contains only 92\% of the flux, compared to a pixel at field center
(see MosManual for details).  Therefore, although an individual star anywhere
on the image will have the same number of photons within the 
point-spread-function (PSF), the variable pixel scale causes the photometric
zero point to vary by 8\% over the field of view.  We correct for this
photometric effect following the recommendations given in the MosManual.
Briefly, we flatten each image with a flat-field frame that has {\it not}
been corrected for the variable pixel scale.  Then, following our astrometric
calibration and prior to stacking multiple exposures, we re-grid each
frame to a tangent-plane projection with pixels of constant angular scale.
We note that during the re-gridding, we {\it do not} scale each
pixel photometrically by the amount each pixel area has changed.  In
this manner we account for the variable pixel scale and produce
uniform images over the entire field of view.

We construct flat-field frames for each passband using a combination of
twilight and night-sky flats.  First we make a normalized flat with high
signal-to-noise (S/N) by averaging a set of twilight illuminated exposures.
This component accounts for both the small scale, high frequency 
(pixel-to-pixel) variations in response, and the spatial variations over
large fractions of an image.  We then fit 
a smooth surface to a night-sky flat produced by median combination of a set of
unflattened frames of the J0053+1234 region with all objects masked out,
and we multiply this smooth surface to the twilight flat frame to produce a
high S/N flat with the spectral response of the night sky.
We iterate the flat field construction twice to optimize the night-sky flat
object masking.  Thus, the resultant ``super flat'' for each passband is
spatially and spectrally flat.
We divide each exposure of the J0053+1234 region and the standard star 
cluster image by its
super flat to achieve $\lesssim 1$\% global flatness over the eight chip array.

Good astrometry is required to register and stack each dithered set of
exposures.  In addition, accurate celestial coordinates are necessary for the
follow-up spectroscopy of identified quasar candidates.
All Mosaic images have an initial default world coordinate system (WCS) loaded
in their header at the time of observation.  The WCS maps the image pixel
space onto celestial coordinate axes (RA and Dec).  However, effects such as
global pointing offsets, instrument rotations and differential atmospheric
refraction produce the need for corrections to this astrometric calibration.
We use the {\it msccmatch} package in IRAF to interactively derive accurate 
(RMS~$\lesssim 0.3\arcsec$) astrometric solutions for each
dithered exposure by matching $\sim300$ fairly bright stars (blue and red
magnitudes $12\lesssim m \lesssim 16.5$), distributed evenly
over the Mosaic field, with a reference frame given by their epoch J2000.0 
USNOv2.0 coordinates \citep{monet96}.  We map the
eight chip exposures for each Mosaic frame onto a single image by rebinning
the pixels to a tangent-plane projection, thus producing pixels of constant
angular size (as described above).  
The resultant image is astrometrically calibrated to the
J2000.0 celestial equatorial WCS. 

Finally, we combine each set of fully processed
and registered exposures into a final high S/N image for each passband.
We subtract a constant flux level equal to the mode sky value
from each image in a dithered set.  Higher order terms are unnecessary due
to the better than 1\% global flat fielding.
Before combining a dithered sequence of exposures into a single image,
we account for the different photometric depths of individual frames.
These differences are due to time-varying effects during observations.  The
most common effect is the changing airmass over an hour-long dithered sequence.
Variable sky transparency due to occasional thin clouds (cirrus) is a second
effect.  All exposures
of a dithered sequence are scaled to the reference image selected to have the
lowest airmass and/or best photometric conditions.  We calculate each frame's
multiplicative scale factor by comparing simple aperture flux measurements
from the set of $\sim300$ astrometric calibration stars common to each
image and its reference.  The scaling factors are typically of order a few
percent.  This procedure scales each set of dithered, same passband frames to
the same effective airmass and exposure time (given in Table \ref{photcal}).
We align and median combine (for removal of cosmic rays) 
each dithered set of registered, scaled,
and sky-subtracted J0053+1234 region Mosaic frames to produce calibrated
and cosmetically clean images.

\subsection{Calibrations and Catalogs}
We use the source detection and extraction software SExtractor 
\citep[Source Extractor;][]{bertin96}
to compile catalogs of instrumental 
magnitudes (MAG\_BEST) for every source in the final $UBRI$ images.
In addition, SExtractor provides accurate positions and a variety of
photometric measurements for each detected source.
We configure SExtractor to
detect objects comprised of a minimum of 5 pixels (DEBLEND\_MINAREA)
above a background threshold of $3\sigma_{\rm bkg}$ (DETECT\_THRESH).  
Overlapping sources are deblended into multiple objects if the contrast
between flux peaks associated with each object is $\geq0.05$ (DEBLEND\_MINCONT).
These parameters provide our
working definition of an imaged source.  We confirm that these parameters
provide good source detection and deblending by visually inspecting
random regions from each image.  We remove sources from each
catalog with saturated, or otherwise
corrupted, pixel values as flagged by SExtractor (FLAGS$\ge4$).  The
primary source of flagged sources are those with bright magnitudes; 
i.e. have at least one saturated pixel.  Additionally,
we exclude sources within 140 pixels ($60\arcsec$) of an image edge; 
stacking and combining dithered frames makes these regions of the 
final images lower in S/N.  

We determine a turn-over magnitude 
($m_{\rm TO}$) where the source counts distribution flattens and begins to
fall off.  This empirical limit provides a rough estimate of the magnitude
that all sources, point-like and extended, become incomplete.  For point
sources we estimate
99\% and 90\% completeness limits by randomly distributing
artificial stars in each image 
(100 per $\Delta m=0.25$ bin over the magnitude
range $m_{\rm TO}-1\leq m \leq m_{\rm TO}+2$), rerunning SExtractor,
and determining the number of these stars that are recovered.  We use the
IRAF {\it artdata} package to create artificial stars with
characteristics (PSF size, magnitude zero point, gain, and Poisson noise) 
matched to actual stars on our images.  We note
that extended source completeness is more difficult to quantify, due
mostly to the wide range of galaxy surface brightnesses.  For this reason,
the completeness of extended sources turns over more slowly than for
point sources.  As can be
seen in Table \ref{cats}, the 90\% completeness limit for point sources
is fainter than $m_{\rm TO}$.

We summarize the magnitude limits and number counts for the sources in the
separate $U$, $B$, $R$, and $I$ catalogs in Table \ref{cats}.
To generate the catalogs from which quasar candidates will be selected, 
we sequentially match the
$U$ catalog with the $B$, $R$, and finally 
the $I$ source catalogs.  We note that in all four cases we use the entire
(i.e. not magnitude-limited) catalogs.
The 1946 $U$ sources are correlated with the $B$ sources resulting
in 1730 $UB$ matches with 216 $U$-only sources (see discussion 
in \S\ref{candsel}).
Comparing the $UB$ matches with the $R$ catalog results in 1642 $UBR$ matches,
leaving 88 $UB$ matches without $R$ magnitudes.  Comparing the $UBR$ matches
with the $I$ catalog results in 1552 $UBRI$ matches, leaving 90 $UBR$ matches
without $I$ magnitudes.  Our final two source catalogs include all
sources with reliable photometry.  They are (1) the $UBRI$
catalog with the 1552 matches in the $UBRI$ bands, and (2) the $UB$ catalog
consisting of the 178 sources with no $I$ magnitudes, of which 88 have
neither $R$ nor $I$ magnitudes.

We transform science image fluxes into apparent magnitudes calibrated to the
\citet{landolt92} system.  Even though the observing conditions were not
photometric, calibrating the photometry to within $\sim25\%$ absolute 
provides
useful flux estimates for the followup spectroscopic observations.
The photometric system is defined by the zero point zp, extinction
(or airmass) coefficient $\alpha$, and color coefficient $\beta$ for each
passband.  These coefficients are determined by solving simple,
linear transformation equations that relate instrumental magnitudes
($u,b,r,i$) of standard stars observed each night with their published
magnitudes ($U,B,R,I$).  We measure the fluxes of standard stars
within a $14\farcs4$ circular aperture
similar to that used by \citet{landolt92}.  
We use IRAF's {\it photcal} package to find the best-fit
solutions to the following transformation equations:
\begin{equation}
U = u + {\rm zp}_U + \alpha_U X_U + \beta_U (U-B)
\end{equation}
\begin{equation}
B = b + {\rm zp}_B + \alpha_B X_B + \beta_B (U-B)
\end{equation}
\begin{equation}
R = r + {\rm zp}_R + \alpha_R X_R + \beta_R (R-I)
\end{equation}
\begin{equation}
I = i + {\rm zp}_I + \alpha_I X_I + \beta_I (R-I)
\end{equation}
We give the calibration coefficients and their errors in Table \ref{photcal}.
The photometric zero point quantifies the gain and
the total sensitivity of the telescope plus detector.  The airmass term
is a measure of the atmospheric extinction as a function of telescope altitude.
The
color term shows how well the instrumental system matches the
\citet{landolt92} system.  We note that
large uncertainties in coefficients ({\it i.e.} systematic zero point and
airmass term errors in excess of 0.1 mag) indicate non-photometric conditions
during the nights we observed standards.

From the instrumental magnitude $m_A$ of each source in passband $A$,
we first convert to apparent magnitude $A=m_A + {\rm zp}_A + \alpha_AX_A$,
using the coefficients given in Table \ref{photcal}.  Next, we calculate the colors
of each source by iteratively solving the following equations:
\begin{equation}
(U-B)_i = (U-B)_0 + (\beta_U - \beta_B)\cdot(U-B)_{i-1} ,
\end{equation}
\begin{equation}
(R-I)_i = (R-I)_0 + (\beta_R - \beta_I)\cdot(R-I)_{i-1} .
\end{equation}
The initial colors $(U-B)_0$ and $(R-I)_0$ are derived simply from the
apparent magnitudes.
We iterate these calculations using color coefficients from Table \ref{photcal}
until
the difference between successive iterations is
$\delta m \leq 0.001$~mag.

Finally, we correct our magnitudes and colors for Galactic extinction by
applying reddening corrections
using the dust maps of \citet{schlegel98}\footnote{The authors have kindly
made available the dust maps with user friendly software that outputs
$E(B-V)$ for an input $(l,b)$ at
http://astron.berkeley.edu/davis/dust/index.html.}.
These new maps are based on
full-sky $100\micron$ emission from {\it COBE}/DIRBE and {\it IRAS}/ISSA
observations and, thus, directly measure the Galactic dust content.
The \citet{schlegel98} data have orders of magnitude
higher resolution (at $6.1\arcsec$), 
and are $\sim2$ times more accurate than the traditional
\citet{burstein82} reddening estimates based on neutral hydrogen $21$~cm
emission.  The J0053+1234 field does not have large extinction;
the local reddening $E(B-V)$ ranges from 0.058 to 0.075, corresponding to
mean extinction corrections of 0.37, 0.29, 0.18, and 0.13~mag for $UBRI$ 
passbands, respectively.
These corrections have a formal uncertainty of 10\%.

The systematic errors of our photometry dominate over random 
errors, even at the faintest magnitudes.
The systematic uncertainties in our photometry are due mainly
to the zero point calibration ($\lesssim0.10$ mag) and the airmass 
correction ($\lesssim0.10$ mag) resulting
from the variable transparency during our observations.  
Yet, the obvious
stellar locus of main sequence stars in the color-color plots (see
\S\ref{candsel})
illustrate the utility of relative
photometry during non-photometric conditions.

\section{Quasar Candidate Selection}
\label{candsel}
The practice of using multi-color photometry to search for quasars is
well known \citep[e.g.][]{koo86,hall96,liu99}.  
In this paper we present the full $UBRI$ photometry, but only use the $U$,
$B$, and $R$ filters for quasar selection
by ultra-violet excess, which is most efficient at $z\lesssim2.5$.
In future work the $I$ information can and will be used to help select
quasars at higher redshift, where the baseline to longer wavelengths
is essential but the efficiency of selection is lower.

\begin{figure}[hp]
  \centering
  \mbox{
\includegraphics[scale=1.1, angle=0]{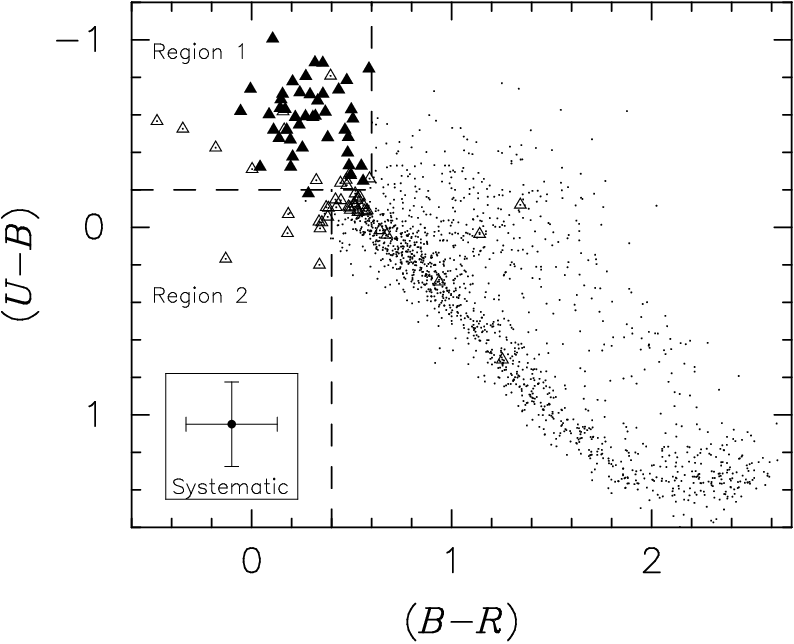}
  }
\caption{
Multi-color photometry of the J0053+1234 field.
$(U-B)$ versus $(B-R)$ color-color plot for the 1552 sources with $UBRI$ 
photometry.  The photometry presented here has been corrected for
airmass and color terms, as well as
Galactic extinction using the \citet{schlegel98} reddening maps.
The average systematic color errors, which dominate over random
photometric errors, are given in the lower left box.  There are 44
spectroscopically confirmed quasars (solid triangles) and 48 confirmed
non-quasi-stellar objects (open triangles).  The boundary (dashed
line) represents the estimated end of the stellar locus; therefore,
sources bluer than this are likely quasars (see text for details).
}
\label{ubr}
\end{figure}

\begin{figure}[hp]
  \centering
  \mbox{
\includegraphics[scale=1.1, angle=0]{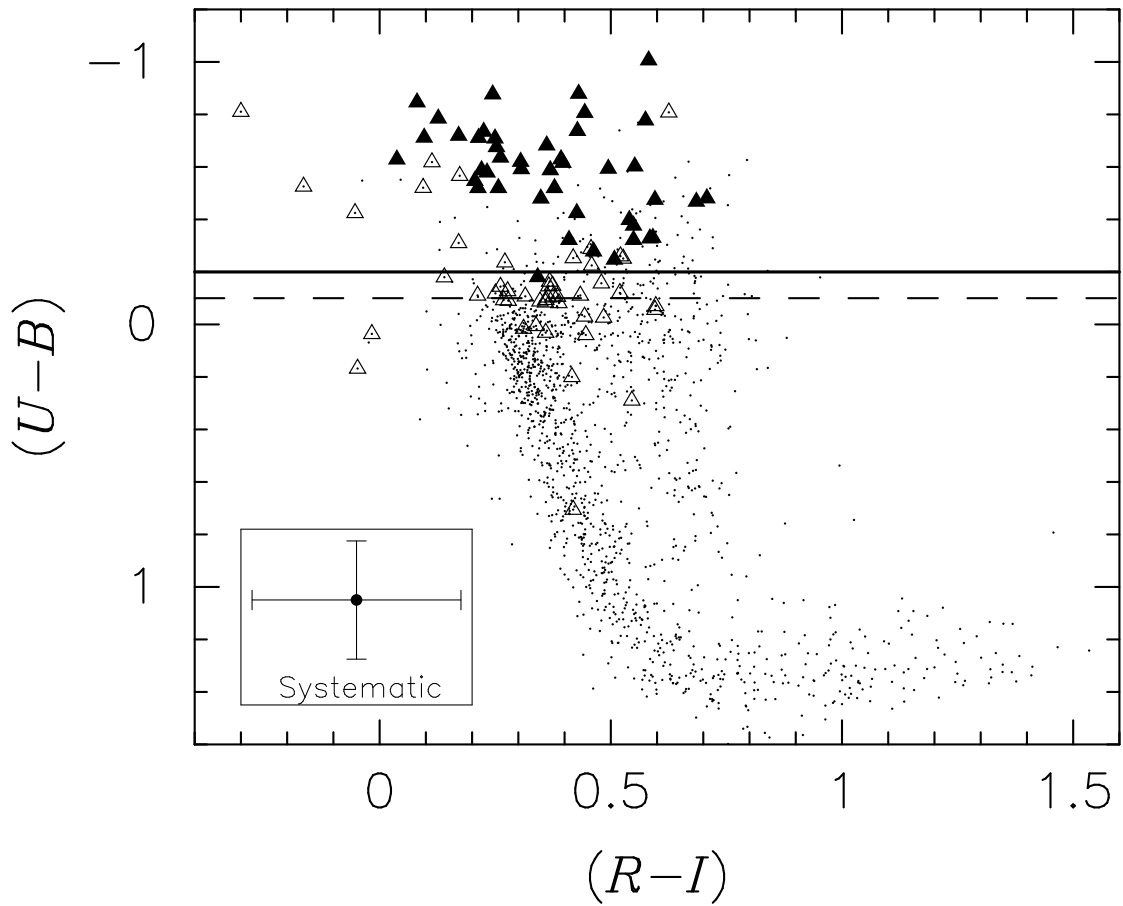}
  }
\caption{
Multi-color photometry of the J0053+1234 field.
$(U-B)$ versus $(R-I)$ color-color plot for the 1552 sources with $UBRI$
photometry.  The photometry presented here has been corrected for
airmass and color terms, as well as
Galactic extinction using the \citet{schlegel98} reddening maps.
The mean systematic color errors, which dominate over random
photometric errors, are given in box at the lower left.  
A cut of $B<21.3$ is applied to keep random photometric errors low enough
that the stellar locus is well-defined.
As in Figure \ref{ubr}, the 44
spectroscopically confirmed quasars (solid triangles) and 48 confirmed
non-quasi-stellar objects (open triangles) are shown.  We show the 
$(U-B)\leq -0.2$ (solid line) and $(U-B)\leq -0.1$ (dashed line) boundaries.
}
\label{ubri}
\end{figure}

\begin{figure}[hp]
  \centering
  \mbox{
\includegraphics[scale=1.1, angle=0]{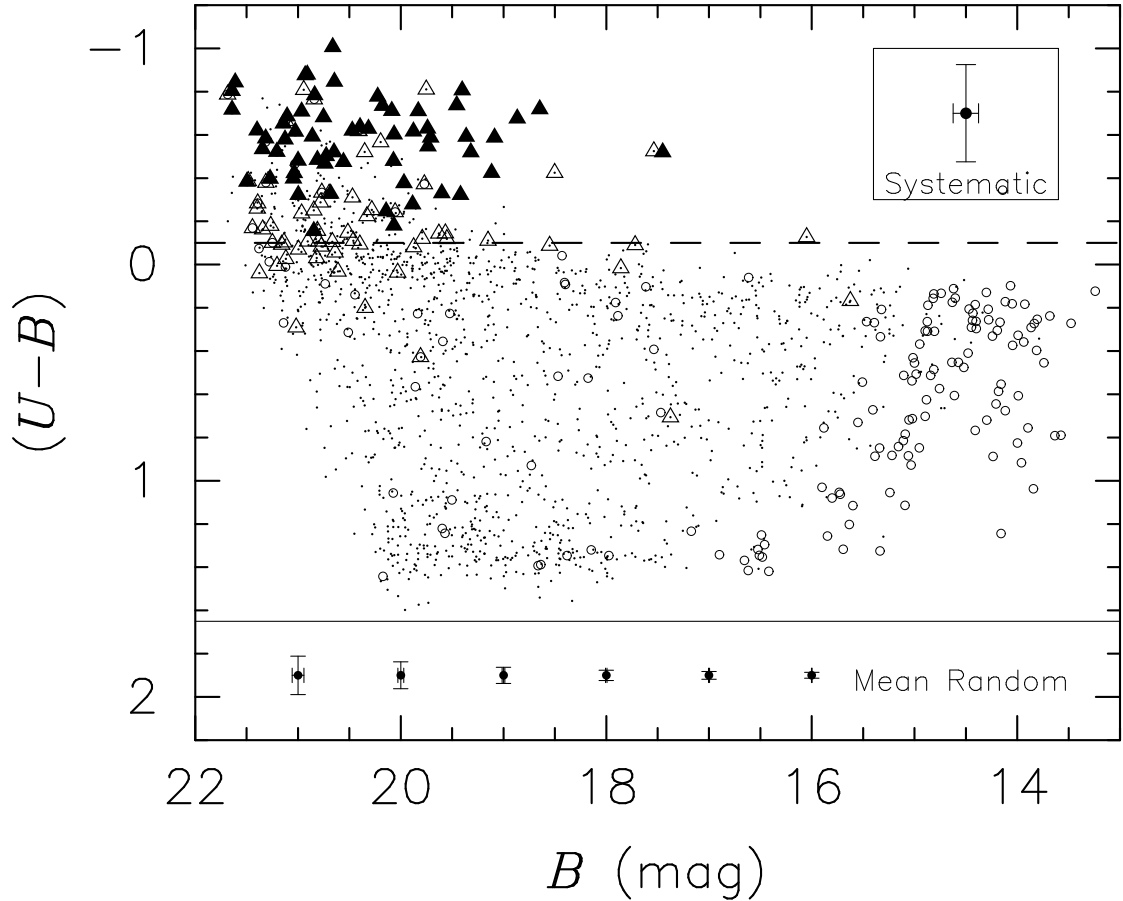}
  }
\caption{
$U,B$ color-magnitude diagram of 1730 sources from the J0053+1234 field,
showing 178 sources with only $U,B$ detections (small open circles), 
and 1552 with $UBRI$ photometry (small dots).  
Sources with spectroscopic observations are
shown as triangles, they comprise 60 confirmed quasars (solid triangles) and
62 nonquasi-stellar objects (open triangles).  One quarter (30/122) of 
the spectroscopic targets are $UB$-only sources, and 16 of these are
confirmed quasars.
The dashed line shows the blue $(U-B)$ color cut for objects with
$R$ and $I$ non-detections.
The photometry presented here has been corrected for
airmass and color terms, as well as
Galactic extinction using the \citet{schlegel98} reddening maps.  We show
the mean systematic error in photometry in the upper right-hand box.  The
random photometric errors as a function of $B$ magnitude are given at the
bottom.
}
\label{ub}
\end{figure}

We plot our photometry for 
the 1552 sources from the $UBRI$ catalog in two color-color plots:
the ($U-B$) vs. ($B-R$) plane in Figure \ref{ubr}; and  ($U-B$) vs. ($R-I$)
in Figure \ref{ubri}.  These sources have an effective cut of
$B=21.3$ mag, corresponding to the limit of photometry accurate enough
to define a tight stellar locus.
In Figure \ref{ub} we plot the ($U-B$) vs. $B$ color-magnitude
diagram for the 1552 sources from the $UBRI$ catalog plus the 178
additional sources from the $UB$ catalog (i.e. only $U$ and $B$ detections).
We note that the majority of sources from the $UB$ 
catalog (plotted in Figure \ref{ub} as open circles) are bright ($B<17$ mag) 
and red ($U-B>0$).  These bright $UB$ sources are saturated in $R$
and $I$, and they
are both too bright and too red to be quasars.  Also, 90 of these
$UB$ sources have $R$ magnitudes but no $I$ magnitudes, and so were not used 
in the ($U-B$) vs. ($B-R$) candidate selection process; however, this means 
that any quasars in this group of 90 will be found at a lower yield.  We
find 34 $UB$ catalog sources scattered between $16.0<B<19.5$ that are not
part of the $UBRI$ catalog for a variety of reasons: (i) 7 are edge
sources in the $R$ or $I$ images\footnote{Our $UBRI$ images are concentric
to within a few tens of pixels ($<10\arcsec$), thus, a slight shift means
some objects will be culled from one catalog given our edge proximity
flag, yet not culled from another.}; (ii) 10 are flagged and removed from
$R$ or $I$ owing to flux contamination from nearby bright red stars
that are not contaminated at bluer passbands; and (iii) 17 are $I$-band
saturated and thus, among the 90 $UBR$ detections.  We note that it
is possible that some of the 88 $UB$ sources that were not detected in $R$
may have been detected in $I$; however, given the completeness limits
of the survey, such objects must have an inflected spectrum 
and are therefore unlikely to be quasars with a power-law
spectral energy distribution.  Lastly, at faint $B>20$ limits there
are blue $(U-B)<0.2$ $UB$ catalog sources that are true $R$ and $I$-band
drop outs, many (30) of which meet our quasar candidate selection (see below)
and were later targeted for spectroscopy (triangles in Figure \ref{ub};
16 are confirmed quasars shown as solid triangles).

Our primary selection strategy is based on ($U-B$) color, and thus depends
on point sources detected jointly in at least $U$ and $B$. However, there 
is an additional category of sources that are potentially of interest 
in a quasar selection experiment: $U$-only detections where the level 
of the $B$ non-detection implies a source blue enough to be a quasar. 
Most of the 216 $U$-only sources are saturated stars or artifacts, and 
53 of the 54 that are plausible point sources are faint, $U > 21$. Thus,
the $U$-only faint sources are fainter than the sky brightness 
at longer wavelengths 
so they were considered too faint for useful spectroscopy. The omission 
of this category, which may include unrecognized quasars, affects the 
overall completeness measures that will be considered in an upcoming 
paper.

As illustrated in Figures \ref{ubr} and \ref{ubri}, the stellar main-sequence 
is readily observed in color-color plots.  The tight locus begins with cool 
stars at $(U-B) \sim 1.4$ and ends abruptly at a blue color of 
$(U-B) \sim -0.2$, corresponding to the hottest main-sequence stars and white 
dwarfs.  Quasars with $z\lesssim2$ have colors typically bluer than the stellar
locus; therefore, following \citet{liu99} we select quasar candidates based on 
$(U-B)$ color, $(B-R)$ color, and $B$ magnitude.  The exact selection procedure
is designed to maximize the efficiency of quasar selection relative to the 
stellar locus.  We select $UBRI$ quasar candidates in the $UBRI$ catalog 
from two regions in the ($U-B$) vs. ($B-R$) plane as shown in Figure \ref{ubr}
by the dashed lines: Region 1 defined by $(U-B)\leq -0.2$ and $(B-R) \leq 0.6$;
and Region 2 defined by $(U-B) > -0.2$ and $(B-R) \leq 0.4$.
In addition, we select the $UB$ quasar candidates
from the $UB$ catalog with $(U-B)\leq -0.1$,
as shown in Figure \ref{ub} by the dashed line. 

We do not attempt to remove extended sources from our photometric catalogs.
Many of the sources redward of the stellar locus in either color 
in Figures \ref{ubr} and 
\ref{ubri} are resolved; nevertheless, their removal to produce a tighter
stellar locus is outweighed by the desire to keep all possible quasar
candidates.  When observed with sufficient resolution and S/N,
many low redshift quasars appear non-stellar because of the presence of a
host galaxy. In principle, any deep efficient search for quasars should
apply a criterion that filters out the substantial number of
faint galaxies while not rejecting quasars whose images are 
softened by host light.  Yet, most of our quasar candidates are
within two magnitudes of the detection limit.  At these apparent
brightnesses,
the large CCD pixel scale and a slightly variable PSF (as a function
of image position) cause SExtractor to be unreliable at
distinguishing stars and galaxies.  Specifically, we find that
both our candidates and our confirmed quasars span the full range of the
SExtractor star/galaxy classifier (CLASS\_STAR).  For these reasons,
we choose not to select according to CLASS\_STAR, and as a result,
our sample has no explicit selection effect against
quasars at low redshift or against quasars with luminous
host galaxies.

We performed several empirical checks of the quality of our
photometry across the large CCD field of view. The surface
density of quasar candidates does not vary significantly
across the field. Nor does the distribution of photometric 
errors, as might occur if there were position-dependent
sensitivity or background variations.  
The variation of random photometric error with $B$ magnitude
is shown in the lower panel of Figure \ref{ub}.
All of this affirms the
homogeneity of the photometry and the resultant catalogs.

\section{Spectroscopy}

\subsection{Observations}
We searched for quasars from the candidate list formed as described in \S2.4
during two observing runs in the Fall of 2000.  For
the first run we used the Hydra multi-object spectrograph on the KPNO 
3.5-meter WIYN 
telescope for 5 nights, 29 Sep -- 3 Oct 2000.  In two overlapping fields
centered on the J0053$+$1234 region, we integrated for 13.5 and 18 hours 
each, observing 62 and 58 targets, respectively.  We combined the Simmons
camera using the the blue fibers with the G400 grating to give wavelength
coverage of about
3410--6610 \AA, with a dispersion of 1.56 \AA\ pix$^{-1}$ 
and 6.8 \AA\ resolution.  We assigned approximately one third of the 100
Hydra fibers to observe the sky for optimum sky subtraction.
During the second run at the MMT 6.5-meter on 30 Nov 2000 UT, we observed 
six targets from the candidate list for 20 min to 1 hour each, depending
on the target brightness.  For these observations, the Blue Channel
spectrograph and the 300 l~mm$^{-1}$ grating (blazed at
5800 \AA) provided wavelength coverage of about 3200--8800 \AA, with a 
dispersion of 1.94 \AA\ pix$^{-1}$ and 8.8 \AA\ resolution.

\subsection{Data Reduction}

The WIYN and MMT data are reduced with standard IRAF routines.  We perform
basic reduction of the CCD data using the {\it imred.ccdred} package.
We trim, subtract the overscan region (fit with a low order Legendre
polynomial), and bias subtract each CCD image. 
The dark current is low enough so that a correction is unnecessary.  
We flat field the MMT data
by fitting a response function in the dispersion direction to a
median-combined dome flat image and dividing it into the science images.
Since only one exposure was taken
of each MMT target, we create a bad pixel mask from the normalized flat field
and use it to replace bad pixels in the science frames with flux values
interpolated from nearby pixels.  

For the WIYN data we perform the spectral extraction and calibration
using the IRAF {\it dohydra} package in {\it imred.hydra}.  We obtained
dome flats for each of the two target configurations, which we use to
trace the positions of the spectra on the chip for extraction as well 
as to determine the throughput correction. The apertures are referenced to the 
fibers so that the extracted spectra are properly identified. 
The flat field correction was performed by fitting an average spectrum over 
all the fibers in a configuration with a high order function. 
For wavelength calibration we calculate a wavelength solution from the
HeNeAr lamp exposures taken adjacent in time to each science exposure.  
The solution
is applied to the quasar candidate, sky, and standard star spectra.  For
each configuration, we visually inspect the set of sky spectra to be sure
each sky position was free from contamination from non-sky sources.
Then we construct an average of good sky spectra
and subtract it from the science and standard star spectra.  We observed
the standard star
BD+28$^\circ$4211 for flux calibration of the science spectra.
This star was observed only once during the run and the conditions were
not photometric, therefore, the spectrophotometric calibration is only accurate
to about 30\% on average.  More accurate spectrophotometry is not required
for quasar identification since the distinguishing features are broad emission 
lines of large equivalent width.

We perform spectral extraction and calibration for the MMT data using
{\it twodspec.longslit}.  We extract
the science and standard star spectra by 
tracing the flux along the dispersion axis and subtracting a value for
the sky. The sky is determined from a linear fit to sky pixels in a region 
transverse to the dispersion direction along the trace.
We observed GD~50 as the flux standard
to calibrate the science spectra.  This
star was only observed once during the run in non-photometric conditions.
Thus, the flux calibration serves primarily to calibrate the continuum shape 
for the science targets, and the flux calibration does not yield 
spectrophotometry.

\section{Results}

\subsection{Quasars}

The number of quasar candidates selected from both the $UBRI$ (Regions 1 and 2)
and $UB$ catalogs by the procedure described in \S\ref{candsel} are listed in 
Table~\ref{qresults}, along with the number of objects that were
observed spectroscopically, and the number of confirmed quasars.  
>From the $UBRI$ selection, Region 1 was the most effective at finding quasars 
with an efficiency of 75\% and a yield of 43 quasars; Region 2 only 
contributed one additional  quasar among 11 targets observed.  From the $UB$ 
selection, 16 quasars were confirmed from 30 targets observed, for an 
efficiency of 53\%. 
Overall, a total of 60
quasars were confirmed with an efficiency of 61\%.
To examine whether a color cut that included the blue end of the stellar
locus would yield more quasars, we observed 12 of 15 sources spectroscopically
with $(U-B)\leq -0.1$ and $0.4<(B-R)\leq 0.6$,  but this selection yielded no
additional quasars.

\begin{figure}[hp]
  \centering
  \mbox{
\includegraphics[scale=0.9, angle=0]{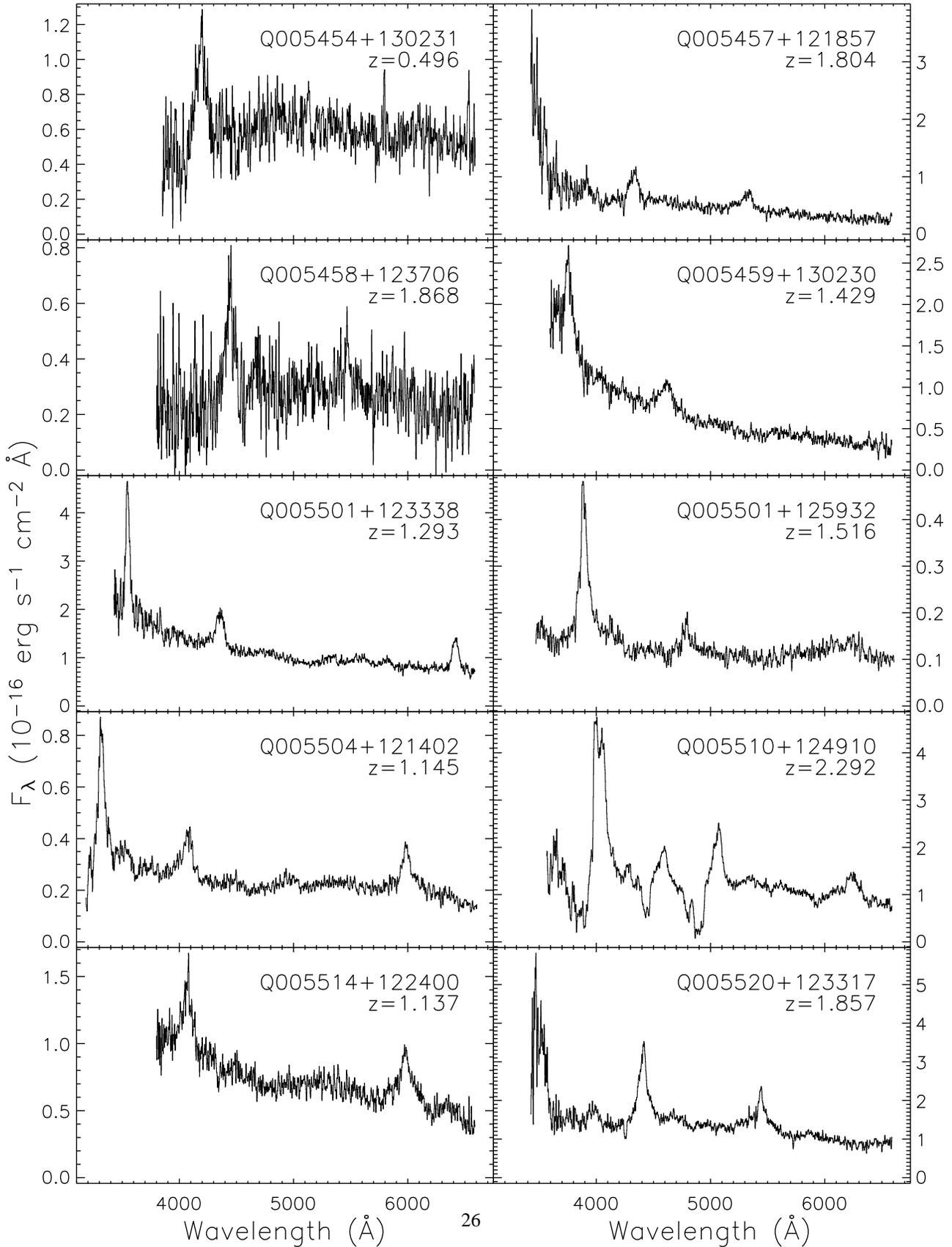}
  }
\vskip -0.5truecm
\caption{
Spectra of 60 newly confirmed quasars plotted in R.A. order from left 
to right and down the page.  Although the spectra have been flux
calibrated, the spectrophotometry is only accurate to about 30\%.
}
\label{qsospec}
\end{figure}

We measured the quasar redshifts and their rms errors
by cross-correlating their spectra with the LBQS composite spectrum 
\citep{fran91} using IRAF's {\it fxcor}. Figure~\ref{qsospec} contains spectra 
of all objects where at least one broad emission line is detected, where we are
essentially certain of the quasar identification. However, redshifts are not
always as certain, particularly when only one emission line is observed. For
spectral coverage in the range 3400--6400 \AA, most redshifts yield two or
more strong lines. The only exception in the range $0.3 < z < 0.8$, where 
Mg~II is the strongest line, C~III] and H$\beta$ are not covered, and [OII]
might not be visible if it is weak. There are two cases (Q~005141$+$123050 
and Q~005344$+$121847) where this occurs, and in each case the single strong
line has only one plausible identification. Several other quasars in 
Figure~\ref{qsospec} 
have their second strong line at the noisy, blue end of the spectrum, but
inspection of the data before clipping for display shows that the lines are
real in each case. We note that three of the 60 spectra in 
Figure~\ref{qsospec} were
taken with the MMT which provides spectral coverage out to 8800 \AA.
In Figure~\ref{qsospec} 
we clip these three spectra to 6750 \AA\ and note that the
only strong line omitted is \ion{Mg}{2} in Q005501+125932, which has
a secure redshift of $z=1.516$.

\begin{figure}[hp]
  \vskip 5truecm
  \centering
  \mbox{
\includegraphics[scale=.7, angle=0]{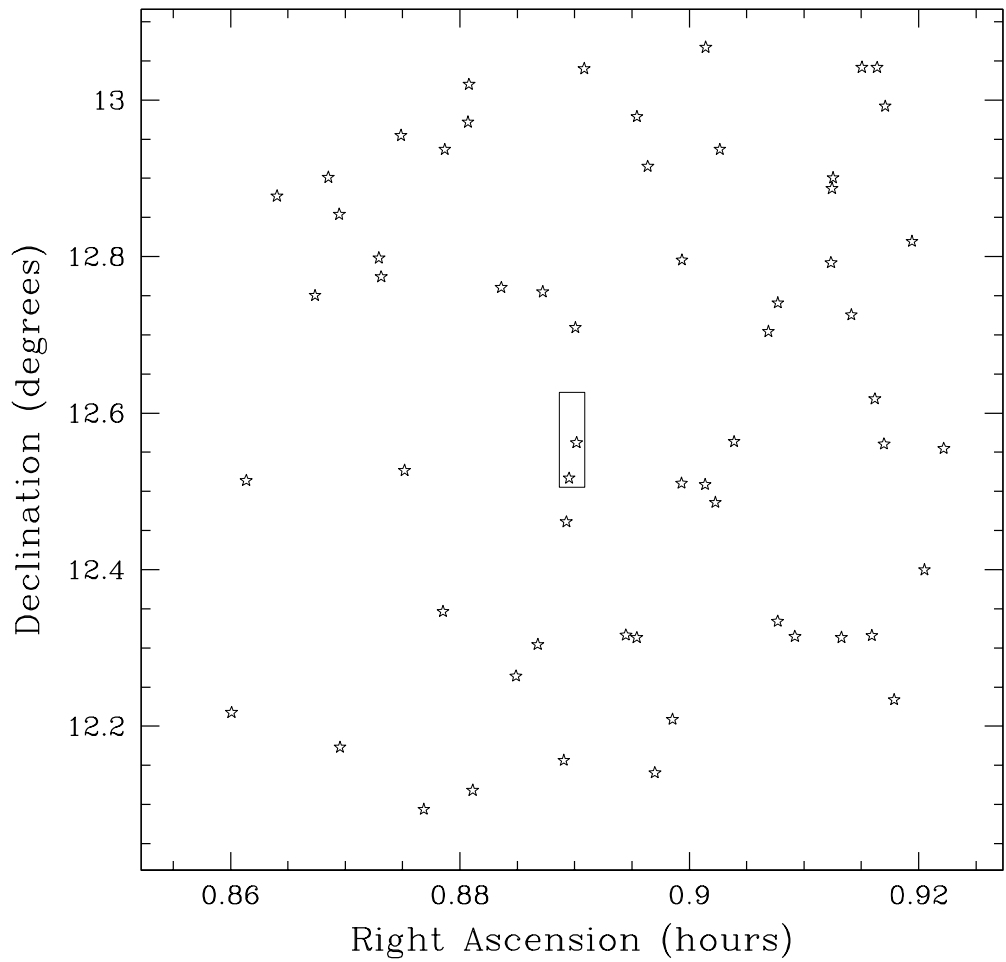}
  }
\caption{
A sky map of the 60 quasars found in this survey.  The rectangle in the 
center of the plot denotes the boundary of the 2\arcmin$\times$ 7\farcm33
\ \citet{cohen99a} galaxy survey.
}
\label{skymap}
\end{figure}

The names of the new quasars, their J2000 coordinates, redshifts, and colors 
are listed in Table \ref{qsos}. Their spectra are plotted in R.A. order in
Figure~\ref{qsospec}.
The spectra have been smoothed with a Gaussian having a FWHM of 3 pixels, and
the pixels at the bluest and reddest ends of the spectra which contain no 
useful information have been trimmed.  Figure~\ref{skymap} shows 
a sky plot of the 60 quasars with the boundaries of the 
galaxy redshift survey of \citet{cohen99b} superimposed.
To see if any quasar had been
discovered previously, we performed a search in NED on the position of each
quasar using a circular radius of 10 arcsec.
Our survey did not rediscover any previously known quasar. NED lists
an X-ray counterpart for Q005321+122740 and a radio counterpart for 
Q005355+121232, but redshifts for neither.

\subsection{Emission Line Galaxies}

\begin{figure}[hp]
  \centering
  \mbox{
\includegraphics[scale=0.9, angle=0]{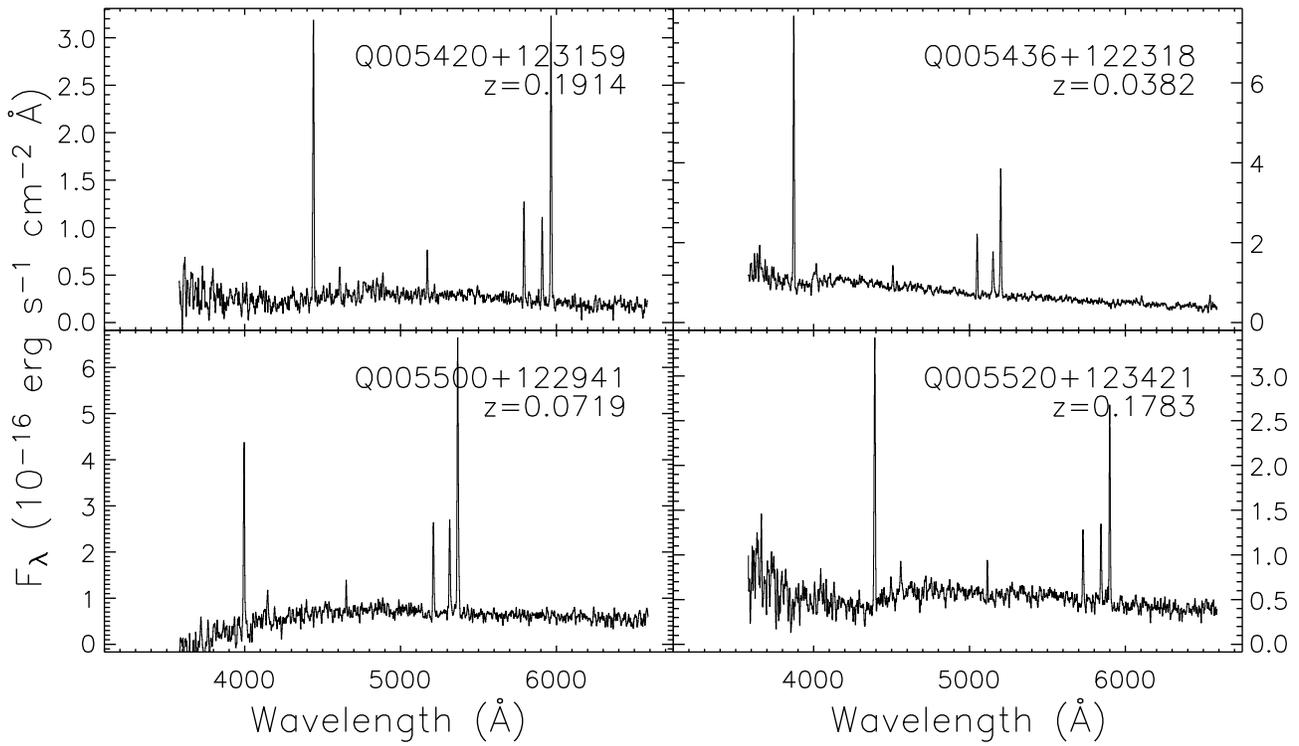}
  }
\vskip -12.0truecm
\caption{
Spectra of 14 new emission line galaxies plotted in R.A. order from left 
to right and down the page.  Although the spectra have been flux
calibrated, the spectrophotometry is only accurate to about 30\%.
}
\label{elgspec}
\end{figure}

Fourteen new narrow emission line galaxies were discovered in this survey.
None were found in a search of NED.
Their redshifts were measured by fitting a Gaussian profile to the
three or four strong lines in each spectrum, computing the redshift
for each line using the fitted line center, and averaging the redshifts
determined for each spectrum.
The ELG names, J2000 coordinates, redshifts, and colors are listed in 
Table~\ref{elgs}. The second lowest redshift galaxy in the Table, 
Q005436+122318 has no colors listed
because the nucleus and a nearby \ion{H}{2} region were separately selected
in the $B$ band and $U$ band images, respectively.
The spectra are plotted in Figure~\ref{elgspec}, 
in R.A. order from left to right and down the page.
The spectra have been smoothed with a Gaussian having a FWHM of 3 pixels, and
the pixels at the bluest and reddest ends of the spectra which contain no
useful information have been trimmed.

\section{Future Work}

We have presented the multi-color photometric selection and 
low-resolution spectroscopic confirmation of 60 new quasars in the 
J0053$+$1234 region.
These sources provide the initial grid of absorption probes
for a cosmological volume centered on this region, which has been the subject
of a deep redshift survey. These probes will be used in future work
involving deep absorption spectroscopy to measure the relationship
between massive halos traced by Mg~II and C~IV absorption, and those traced
by luminous galaxies.  Moreover, absorption in the spectra of these quasars 
will allow us to
measure the bias between the baryons in the massive halos of luminous
galaxies, and the baryons in the intergalactic medium traced by the
Lyman-$\alpha$ forest of HI absorption.  We have begun the deep absorption
spectra observing campaign.  Furthermore, we have
selected additional quasar candidates based on deeper $U$-band imaging of
the central half square degree of this region. 
The goal of the deeper imaging is to have a denser grid of absorption
probes at small impact parameters from the field center, but spectroscopy
of these additional quasars will be very challenging.  Upcoming papers will
address the overall completeness of our survey towards J0053$+$1234, and 
present additional quasar candidates and spectra.

\acknowledgments


\begin{deluxetable}{lcrcccc}
\tablewidth{0pt}
\tablenum{1}
\tabletypesize{\small}
\tablecolumns{7}
\tablecaption{Journal of Observations for the J0053+1234 Region}
\tablehead{\colhead{UT Date} & \colhead{Filter} & \colhead{$t_{\rm exp}$} & \colhead{N$_{exp}$} & \colhead{Seeing} & \colhead{Airmass} & \colhead{Conditions} \\
\colhead{(1)} & \colhead{(2)} & \colhead{(3)} & \colhead{(4)} & \colhead{(5)} & \colhead{(6)} & \colhead{(7)} }
\startdata
30 Sep 1998 & $U$ &  175 & 5 & 1.7 & $1.06-1.37$ & Mostly clear \\ 
            & $B$ &  30 & 2 & 2.0 & $1.06-1.07$ &  \\ 
            & $B$ &  30 & 3 & 1.9 & $1.13-1.16$ & \\ 
            & $R$ &  30  & 5 & 1.9 & $1.22-1.33$ & \\ 
            & $I$ & 10  & 1 & 1.7 & $1.08$ &  \\ 
01 Oct 1998 & $U$ & 150  & 5 & 1.9 & $1.09-1.34$ & Mostly clear \\ 
            & $I$ & 60  & 5 & 1.6 & $1.40-1.19$ & \\ 
02 Oct 1998 & $U$ & 125  & 5 & 1.8 & $1.12-1.48$ & Some thin cirrus \\ 
            & $B$ & 50  & 5 & 1.6 & $1.06-1.07$ & \\ 
\enddata
\tablecomments{Date of observation (1) and filter employed (2).  The total 
exposure time in minutes (3), and the number of exposures for a dithered 
sequence of images (4).  The estimated mean seeing FWHM in arcseconds (5),
range of airmass values during each exposure sequence (6), and the
conditions during the observations (7).}
\label{obslog}
\end{deluxetable}

\clearpage

\begin{deluxetable}{lcccccccc}
\tablewidth{0pt}
\tablenum{2}
\tabletypesize{\small}
\tablecolumns{9}
\tablecaption{Photometric Calibration Parameters}
\tablehead{\colhead{Filter} & \colhead{$\tilde{t}_{\rm exp}$} & \colhead{$\tilde{X}$} & \colhead{Sky} & \colhead{UT Date} & \colhead{zp} & \colhead{$\alpha$} & \colhead{$\beta$} & \colhead{RMS} \\
\colhead{(1)} & \colhead{(2)} & \colhead{(3)} & \colhead{(4)} & \colhead{(5)} & \colhead{(6)} & \colhead{(7)} & \colhead{(8)} & \colhead{(9)} }
\startdata
$U$ & 35 & 1.06 &   97 & 30 Sep 1998 & $20.00\pm0.12$ & $-0.51\pm0.09$ & $+0.04\pm0.03$ & 0.051 \\ 
$B$ & 10 & 1.06 &  462 & 30 Sep 1998 & $21.70\pm0.07$ & $-0.11\pm0.05$ & $+0.12\pm0.01$ & 0.029 \\
$R$ & 6  & 1.21 &  449 & 30 Sep 1998 & $22.33\pm0.14$ & $-0.33\pm0.10$ & $+0.06\pm0.03$ & 0.036 \\
$I$ & 12 & 1.19 & 1328 & 1 Oct 1998 & $21.37\pm0.09$ & $+0.01\pm0.06$ & $+0.01\pm0.02$ & 0.030 \\
\enddata
\tablecomments{The combined image parameters for each passband 
filter (1) are given: the effective image exposure time in 
minutes (2), the effective image airmass (3), and the scaled average of the
background modes in ADUs (4).
The transformation variables for photometric calibration are given
for each filter: the UT date the photometric standards were observed (5),
the magnitude zero point (6), the airmass coefficient (7), and the color
term coefficient (8).  The RMS in magnitudes (9) 
is given for the best-fit to the transformation
equation.}
\label{photcal}
\end{deluxetable}

\clearpage

\begin{deluxetable}{lccccccccc}
\tablewidth{0pt}
\tablenum{3}
\tabletypesize{\small}
\tablecolumns{10}
\tablecaption{$UBRI$ Source Catalogs}
\tablehead{\colhead{Passband} & \colhead{} & \multicolumn{3}{c}{$99\%$ Completeness Limit} & \multicolumn{3}{c}{$90\%$ Completeness Limit} & \colhead{} & \colhead{} \\
\colhead{} & \colhead{$N_{\rm tot}$} & \colhead{$m$} & \colhead{$N$} & \colhead{S/N} & \colhead{$m$} & \colhead{$N$} & \colhead{S/N} & \colhead{$m_{\rm TO}$} &\colhead{Seeing} \\
\colhead{(1)} & \colhead{(2)} & \colhead{(3)} & \colhead{(4)} & \colhead{(5)} & \colhead{(6)} & \colhead{(7)} & \colhead{(8)} & \colhead{(9)} & \colhead{(10)} }
\startdata
$U$ & 1946 & 20.7 & 1283 & $\sim25$ & 21.5 & 1754 & $\sim12$ & 21.0 & 1.7 \\ 
$B$ & 2463 & 20.7 & 1575 & $\sim25$ & 21.7 & 2334 & $\sim11$ & 21.2 & 1.8 \\ 
$R$ & 4505 & 19.5 & 2080 & $\sim33$ & 20.5 & 3678 & $\sim14$ & 20.2 & 1.9 \\
$I$ & 7922 & 19.5 & 3493 & $\sim25$ & 20.5 & 6569 & $\sim11$ & 20.2 & 1.7 \\
\enddata
\tablecomments{For each SExtractor generated catalog of passband (1), 
the total number of sources 
(2) following the removal of saturated and edge proximity objects (see text).
The $99\%$ point-source
completeness limit magnitude (3), number of sources (4), and signal-to-noise
limit (5).  (6-8) give similar values for $90\%$ 
point-source completeness.  The magnitude
at which source number counts turn over from steeply rising is in (9).
The mean seeing FWHM in arcseconds for each combined image is in (10).}
\label{cats}
\end{deluxetable}

\clearpage

\begin{deluxetable}{lrrrr}
\tablewidth{0pt}
\tablenum{4}
\tabletypesize{\small}
\tablecolumns{5}
\tablecaption{Quasar Survey Yield}
\tablehead{
\colhead{Catalog} & 
\colhead{Number of} & 
\colhead{Number} & 
\colhead{Number of} & 
\colhead{Yield} \\
\colhead{} & 
\colhead{Candidates} & 
\colhead{Observed} & 
\colhead{Confirmed Quasars} &
\colhead{} \\
\colhead{(1)} & 
\colhead{(2)} & 
\colhead{(3)} & 
\colhead{(4)} &
\colhead{(5)} 
}
\startdata
UBRI -- Region 1\tablenotemark{a}  & 58~~~~~ & 57~~~~~ & 43~~~~~~~~~~~~ & 75\%  \\
UBRI -- Region 2\tablenotemark{b}  & 15~~~~~ & 11~~~~~ & 1~~~~~~~~~~~~  & 9\%  \\
UB\tablenotemark{c}                & 32~~~~~ & 30~~~~~ & 16~~~~~~~~~~~~ & 53\%  \\
Combined          & 105~~~~~ & 98~~~~~ & 60~~~~~~~~~~~~ & 61\%  \\
\enddata
\tablenotetext{a}{Objects with $(U-B)\leq -0.2$ and $(B-R) \leq 0.6$; Region 1 
in Figure \ref{ubr}.}
\tablenotetext{b}{Objects with $(U-B) > -0.2$ and $(B-R) \leq 0.4$; Region 2
in Figure \ref{ubr}.}
\tablenotetext{c}{Objects with $(U-B)\leq -0.1$; see Figure \ref{ub}.}
\tablecomments{ 
For each color selection criterion (1), the number of objects meeting that
criterion (2), the number of objects actually observed (3), the number of
quasars found (4), and the efficiency of finding quasars (5) for that 
criterion.
}
\label{qresults}
\end{deluxetable}

\clearpage

\begin{deluxetable}{lccccrrr}
\tablewidth{0pt}
\tablenum{5}
\tabletypesize{\footnotesize}
\tablecolumns{8}
\tablecaption{Quasars in the J0053+1234 Region}
\tablehead{\colhead{Name} & \colhead{RA$_{2000}$} & \colhead{Dec$_{2000}$} & \colhead{$z$} & \colhead{$B$} & \colhead{$(U-B)$} & \colhead{$(B-R)$} & \colhead{$(R-I)$} \\
\colhead{(1)} &
\colhead{(2)} &
\colhead{(3)} &
\colhead{(4)} &
\colhead{(5)} &
\colhead{(6)} &
\colhead{(7)} &
\colhead{(8)} 
}
\startdata
Q005127+121931 & 00 51 26.94 & +12 19 30.5 & $1.027 \pm 0.007$ & $20.65\pm 0.04$ & $-0.52\pm 0.05$ & $ 0.47\pm 0.07$ & $ 0.38\pm 0.08$ \\
Q005136+121303 & 00 51 36.29 & +12 13 02.9 & $1.735 \pm 0.003$ & $19.09\pm 0.01$ & $-0.59\pm 0.02$ & $ 0.22\pm 0.03$ & $ 0.37\pm 0.03$ \\
Q005141+123050 & 00 51 40.90 & +12 30 49.9 & $0.686 \pm 0.011$ & $19.97\pm 0.03$ & $-0.38\pm 0.04$ & $ 0.20\pm 0.06$ & $ 0.55\pm 0.06$ \\
Q005151+125239 & 00 51 50.57 & +12 52 38.5 & $2.076 \pm 0.001$ & $20.66\pm 0.06$ & $-1.01\pm 0.07$ & $ 0.10\pm 0.10$ & $ 0.58\pm 0.10$ \\
Q005202+124501 & 00 52 02.50 & +12 45 01.1 & $2.119 \pm 0.001$ & $21.11\pm 0.08$ & $-0.69\pm 0.10$ & \nodata & \nodata \\
Q005207+125404 & 00 52 06.68 & +12 54 04.4 & $1.200 \pm 0.010$ & $20.75\pm 0.06$ & $-0.68\pm 0.07$ & $ 0.15\pm 0.10$ & $ 0.36\pm 0.10$ \\
Q005210+125114 & 00 52 10.10 & +12 51 14.1 & $2.127 \pm 0.001$ & $21.14\pm 0.08$ & $-0.65\pm 0.10$ & \nodata & \nodata \\
Q005210+121023 & 00 52 10.36 & +12 10 23.0 & $1.656 \pm 0.006$ & $21.04\pm 0.07$ & $-0.42\pm 0.09$ & \nodata & \nodata \\
Q005223+124754 & 00 52 22.57 & +12 47 53.6 & $1.447 \pm 0.007$ & $21.03\pm 0.08$ & $-0.62\pm 0.10$ & \nodata & \nodata \\
Q005223+124627 & 00 52 23.28 & +12 46 26.9 & $2.386 \pm 0.001$ & $20.85\pm 0.06$ & $-0.16\pm 0.09$ & \nodata & \nodata \\
Q005229+125717 & 00 52 29.49 & +12 57 16.7 & $1.435 \pm 0.008$ & $19.88\pm 0.03$ & $-0.62\pm 0.04$ & $ 0.37\pm 0.05$ & $ 0.40\pm 0.05$ \\
Q005231+123137 & 00 52 30.61 & +12 31 36.5 & $1.311 \pm 0.006$ & $18.87\pm 0.01$ & $-0.68\pm 0.02$ & $ 0.33\pm 0.02$ & $ 0.25\pm 0.02$ \\
Q005237+120538 & 00 52 36.65 & +12 05 37.6 & $0.801 \pm 0.009$ & $20.74\pm 0.07$ & $-0.47\pm 0.09$ & $ 0.19\pm 0.10$ & $ 0.69\pm 0.10$ \\
Q005243+122048 & 00 52 42.68 & +12 20 47.7 & $1.255 \pm 0.009$ & $20.84\pm 0.06$ & $-0.78\pm 0.08$ & $ 0.48\pm 0.09$ & $ 0.13\pm 0.09$ \\
Q005243+125613 & 00 52 43.23 & +12 56 12.6 & $0.809 \pm 0.009$ & $17.45\pm 0.01$ & $-0.52\pm 0.01$ & $ 0.18\pm 0.01$ & $ 0.21\pm 0.01$ \\
Q005251+125819 & 00 52 50.51 & +12 58 18.6 & $0.954 \pm 0.008$ & $18.65\pm 0.01$ & $-0.72\pm 0.02$ & $ 0.24\pm 0.02$ & $ 0.17\pm 0.02$ \\
Q005251+130112 & 00 52 50.83 & +13 01 12.2 & $1.904 \pm 0.004$ & $19.40\pm 0.02$ & $-0.81\pm 0.03$ & $ 0.27\pm 0.04$ & $ 0.44\pm 0.04$ \\
Q005252+120705 & 00 52 52.01 & +12 07 05.4 & $0.638 \pm 0.009$ & $21.05\pm 0.08$ & $-0.40\pm 0.10$ & $ 0.48\pm 0.11$ & $ 0.54\pm 0.10$ \\
Q005301+124538 & 00 53 00.95 & +12 45 37.6 & $0.872 \pm 0.007$ & $21.27\pm 0.08$ & $-0.40\pm 0.11$ & \nodata & \nodata \\
Q005306+121551 & 00 53 05.59 & +12 15 50.8 & $2.318 \pm 0.003$ & $21.50\pm 0.09$ & $-0.38\pm 0.12$ & \nodata & \nodata \\
Q005312+121816 & 00 53 12.42 & +12 18 15.7 & $1.662 \pm 0.007$ & $20.56\pm 0.05$ & $-0.48\pm 0.06$ & $ 0.14\pm 0.08$ & $ 0.60\pm 0.08$ \\
Q005314+124518 & 00 53 13.99 & +12 45 18.2 & $1.631 \pm 0.007$ & $20.68\pm 0.05$ & $-0.33\pm 0.07$ & $ 0.55\pm 0.08$ & $ 0.58\pm 0.07$ \\
Q005321+120923 & 00 53 20.61 & +12 09 23.0 & $0.805 \pm 0.008$ & $21.35\pm 0.09$ & $-0.53\pm 0.11$ & \nodata & \nodata \\
Q005321+122740 & 00 53 21.41 & +12 27 40.3 & $0.546 \pm 0.006$ & $19.12\pm 0.01$ & $-0.43\pm 0.02$ & $ 0.25\pm 0.03$ & $ 0.43\pm 0.03$ \\
Q005322+123101 & 00 53 22.26 & +12 31 00.8 & $1.115 \pm 0.010$ & $21.61\pm 0.10$ & $-0.84\pm 0.12$ & \nodata & \nodata \\
Q005324+124233 & 00 53 24.24 & +12 42 33.2 & $2.146 \pm 0.001$ & $20.07\pm 0.03$ & $-0.60\pm 0.04$ & $ 0.09\pm 0.06$ & $ 0.55\pm 0.06$ \\
Q005325+123344 & 00 53 24.59 & +12 33 44.0 & $0.679 \pm 0.009$ & $19.89\pm 0.03$ & $-0.28\pm 0.04$ & $ 0.50\pm 0.05$ & $ 0.46\pm 0.04$ \\
Q005327+130224 & 00 53 27.00 & +13 02 24.2 & $1.182 \pm 0.010$ & $19.83\pm 0.03$ & $-0.71\pm 0.04$ & $ 0.15\pm 0.05$ & $ 0.21\pm 0.05$ \\
Q005340+121859 & 00 53 40.09 & +12 18 59.2 & $0.771 \pm 0.010$ & $19.32\pm 0.02$ & $-0.52\pm 0.02$ & $ 0.11\pm 0.03$ & $ 0.26\pm 0.03$ \\
Q005344+121847 & 00 53 43.51 & +12 18 47.4 & $0.622 \pm 0.010$ & $19.42\pm 0.02$ & $-0.32\pm 0.02$ & $ 0.04\pm 0.03$ & $ 0.41\pm 0.04$ \\
Q005344+125844 & 00 53 43.54 & +12 58 43.8 & $0.655 \pm 0.007$ & $20.14\pm 0.04$ & $-0.25\pm 0.05$ & $ 0.56\pm 0.06$ & $ 0.51\pm 0.05$ \\
Q005347+125455 & 00 53 46.95 & +12 54 54.7 & $0.592 \pm 0.011$ & $19.60\pm 0.03$ & $-0.33\pm 0.03$ & $ 0.49\pm 0.04$ & $ 0.59\pm 0.04$ \\
Q005349+120826 & 00 53 49.19 & +12 08 26.1 & $1.332 \pm 0.006$ & $20.93\pm 0.05$ & $-0.88\pm 0.07$ & $ 0.36\pm 0.09$ & $ 0.24\pm 0.10$ \\
Q005355+121232 & 00 53 54.71 & +12 12 31.5 & $2.089 \pm 0.001$ & $21.64\pm 0.10$ & $-0.80\pm 0.13$ & \nodata & \nodata \\
Q005358+123038 & 00 53 57.57 & +12 30 37.9 & $0.804 \pm 0.009$ & $21.31\pm 0.08$ & $-0.58\pm 0.10$ & \nodata & \nodata \\
Q005358+124744 & 00 53 57.69 & +12 47 44.4 & $2.103 \pm 0.001$ & $20.73\pm 0.07$ & $-0.50\pm 0.09$ & \nodata & \nodata \\
Q005405+123031 & 00 54 04.96 & +12 30 31.3 & $1.574 \pm 0.005$ & $20.07\pm 0.03$ & $-0.48\pm 0.04$ & $ 0.38\pm 0.04$ & $ 0.35\pm 0.04$ \\
Q005405+130403 & 00 54 05.14 & +13 04 02.6 & $1.824 \pm 0.006$ & $19.46\pm 0.02$ & $-0.74\pm 0.03$ & $-0.01\pm 0.04$ & $ 0.43\pm 0.04$ \\
Q005408+122909 & 00 54 08.19 & +12 29 09.3 & $1.162 \pm 0.010$ & $20.65\pm 0.04$ & $-0.85\pm 0.06$ & $ 0.59\pm 0.07$ & $ 0.08\pm 0.07$ \\
Q005410+125613 & 00 54 09.58 & +12 56 13.1 & $1.368 \pm 0.009$ & $20.47\pm 0.05$ & $-0.62\pm 0.06$ & $-0.06\pm 0.09$ & $ 0.31\pm 0.10$ \\
Q005414+123349 & 00 54 14.05 & +12 33 48.9 & $2.098 \pm 0.001$ & $19.36\pm 0.02$ & $-0.59\pm 0.02$ & $ 0.27\pm 0.03$ & $ 0.31\pm 0.03$ \\
Q005425+124215 & 00 54 24.81 & +12 42 15.5 & $1.132 \pm 0.010$ & $19.71\pm 0.02$ & $-0.59\pm 0.03$ & $ 0.31\pm 0.04$ & $ 0.22\pm 0.04$ \\
Q005428+122003 & 00 54 27.74 & +12 20 02.6 & $1.725 \pm 0.005$ & $21.21\pm 0.07$ & $-0.52\pm 0.10$ & \nodata & \nodata \\
Q005428+124427 & 00 54 27.79 & +12 44 27.1 & $2.200 \pm 0.001$ & $20.81\pm 0.06$ & $-0.48\pm 0.08$ & \nodata & \nodata \\
Q005433+121853 & 00 54 33.15 & +12 18 53.1 & $1.699 \pm 0.005$ & $21.00\pm 0.07$ & $-0.32\pm 0.09$ & $ 0.20\pm 0.12$ & $ 0.55\pm 0.11$ \\
Q005444+124732 & 00 54 44.47 & +12 47 31.8 & $1.286 \pm 0.007$ & $20.91\pm 0.07$ & $-0.88\pm 0.08$ & $ 0.32\pm 0.11$ & $ 0.43\pm 0.11$ \\
Q005445+125313 & 00 54 44.79 & +12 53 13.3 & $2.124 \pm 0.001$ & $20.23\pm 0.04$ & $-0.78\pm 0.05$ & $ 0.20\pm 0.07$ & $ 0.57\pm 0.06$ \\
Q005445+125402 & 00 54 45.15 & +12 54 02.0 & $0.868 \pm 0.011$ & $19.74\pm 0.03$ & $-0.55\pm 0.04$ & $ 0.24\pm 0.05$ & $ 0.21\pm 0.05$ \\
Q005448+121848 & 00 54 47.75 & +12 18 47.7 & $0.924 \pm 0.007$ & $21.00\pm 0.07$ & $-0.48\pm 0.09$ & $ 0.48\pm 0.11$ & $ 0.71\pm 0.10$ \\
Q005451+124332 & 00 54 50.86 & +12 43 32.1 & $1.281 \pm 0.009$ & $21.13\pm 0.08$ & $-0.58\pm 0.09$ & $ 0.51\pm 0.12$ & $ 0.23\pm 0.11$ \\
Q005454+130231 & 00 54 54.18 & +13 02 30.6 & $0.496 \pm 0.012$ & $21.40\pm 0.09$ & $-0.62\pm 0.12$ & \nodata & \nodata \\
Q005457+121857 & 00 54 57.30 & +12 18 57.0 & $1.804 \pm 0.005$ & $20.86\pm 0.06$ & $-0.59\pm 0.07$ & $ 0.32\pm 0.10$ & $ 0.49\pm 0.10$ \\
Q005458+123706 & 00 54 58.23 & +12 37 05.5 & $1.868 \pm 0.005$ & $21.64\pm 0.10$ & $-0.72\pm 0.12$ & \nodata & \nodata \\
Q005459+130230 & 00 54 58.93 & +13 02 29.6 & $1.429 \pm 0.007$ & $20.97\pm 0.08$ & $-0.71\pm 0.09$ & $ 0.29\pm 0.11$ & $ 0.25\pm 0.12$ \\
Q005501+123338 & 00 55 01.12 & +12 33 38.1 & $1.293 \pm 0.006$ & $20.31\pm 0.04$ & $-0.63\pm 0.05$ & $ 0.50\pm 0.06$ & $ 0.04\pm 0.06$ \\
Q005501+125932 & 00 55 01.48 & +12 59 31.8 & $1.516 \pm 0.006$ & $20.18\pm 0.04$ & $-0.73\pm 0.05$ & $ 0.44\pm 0.06$ & $ 0.22\pm 0.06$ \\
Q005504+121402 & 00 55 04.29 & +12 14 02.1 & $1.145 \pm 0.008$ & $20.09\pm 0.03$ & $-0.71\pm 0.04$ & $ 0.36\pm 0.04$ & $ 0.10\pm 0.05$ \\
Q005510+124910 & 00 55 09.89 & +12 49 09.8 & $2.292 \pm 0.002$ & $20.07\pm 0.03$ & $-0.18\pm 0.05$ & $ 0.28\pm 0.05$ & $ 0.34\pm 0.05$ \\
Q005514+122400 & 00 55 13.82 & +12 23 59.7 & $1.137 \pm 0.010$ & $20.40\pm 0.05$ & $-0.64\pm 0.06$ & $ 0.14\pm 0.07$ & $ 0.26\pm 0.08$ \\
Q005520+123317 & 00 55 19.87 & +12 33 17.4 & $1.857 \pm 0.005$ & $19.74\pm 0.03$ & $-0.63\pm 0.04$ & $ 0.17\pm 0.05$ & $ 0.39\pm 0.05$ \\
\enddata
\tablecomments{
For each new confirmed quasar (1), the J2000 coordinates (2) and (3), the
measured redshift and error (4), the $B$ band brightness in magnitudes (5),
and the measured $U-B$ (6), $B-R$ (7), and $R-I$ (8) colors in magnitudes.
}
\label{qsos}
\end{deluxetable}

\begin{deluxetable}{lccccrrr}
\tablewidth{0pt}
\tablenum{6}
\tabletypesize{\footnotesize}
\tablecolumns{8}
\tablecaption{Emission Line Galaxies in the J0053+1234 Region}
\tablehead{\colhead{Name} & \colhead{RA$_{2000}$} & \colhead{Dec$_{2000}$} & 
\colhead{$z$} & \colhead{$B$} & \colhead{$(U-B)$} & \colhead{$(B-R)$} & 
\colhead{$(R-I)$} \\
\colhead{(1)} &
\colhead{(2)} &
\colhead{(3)} &
\colhead{(4)} &
\colhead{(5)} &
\colhead{(6)} &
\colhead{(7)} &
\colhead{(8)}
}
\startdata
Q005212+120851 & 00 52 11.98 & +12 08 50.8 & $0.1142\pm0.0005$ & $21.00\pm 0.08$ & $-0.07\pm 0.11$ & $ 0.18\pm 0.13$ & $ 0.60\pm 0.12$ \\
Q005247+130321 & 00 52 47.20 & +13 03 20.9 & $0.1906\pm0.0004$ & $20.95\pm 0.07$ & $-0.81\pm 0.09$ & $ 0.39\pm 0.11$ & $ 0.63\pm 0.11$ \\
Q005249+122433 & 00 52 48.71 & +12 24 33.1 & $0.1914\pm0.0005$ & $21.39\pm 0.08$ & $-0.28\pm 0.12$ & \nodata & \nodata \\
Q005255+125842 & 00 52 54.75 & +12 58 42.3 & $0.0788\pm0.0005$ & $20.28\pm 0.06$ & $-0.25\pm 0.08$ & $ 0.47\pm 0.08$ & $ 0.42\pm 0.08$ \\
Q005323+121946 & 00 53 23.22 & +12 19 45.9 & $0.1031\pm0.0005$ & $21.69\pm 0.11$ & $-0.79\pm 0.14$ & \nodata & \nodata \\
Q005334+123353 & 00 53 33.59 & +12 33 52.9 & $0.0576\pm0.0004$ & $20.77\pm 0.06$ & $-0.33\pm 0.10$ & \nodata & \nodata \\
Q005336+125611 & 00 53 36.25 & +12 56 10.6 & $0.0305\pm0.0006$ & $20.46\pm 0.06$ & $-0.11\pm 0.08$ & $ 0.48\pm 0.09$ & $ 0.43\pm 0.09$ \\
Q005339+125158 & 00 53 38.87 & +12 51 57.9 & $0.1132\pm0.0005$ & $20.85\pm 0.07$ & $-0.25\pm 0.10$ & $ 0.32\pm 0.11$ & $ 0.53\pm 0.10$ \\
Q005347+124450 & 00 53 46.96 & +12 44 49.9 & $0.0796\pm0.0004$ & $19.88\pm 0.04$ & $-0.08\pm 0.06$ & $ 0.53\pm 0.06$ & $ 0.39\pm 0.06$ \\
Q005406+125338 & 00 54 06.44 & +12 53 38.0 & $0.3148\pm0.0103$ & $20.70\pm 0.07$ & $-0.33\pm 0.10$ & \nodata & \nodata \\
Q005420+123159 & 00 54 20.33 & +12 31 59.2 & $0.1914\pm0.0005$ & $21.40\pm 0.09$ & $-0.26\pm 0.12$ & $ 0.59\pm 0.13$ & $ 0.52\pm 0.11$ \\
Q005436+122318 & 00 54 36.47 & +12 23 17.8 & $0.0382\pm0.0018$ & \nodata  & \nodata  & \nodata  & \nodata \\
Q005500+122941 & 00 55 00.25 & +12 29 40.7 & $0.0719\pm0.0006$ & $20.64\pm 0.05$ & $-0.06\pm 0.08$ & $ 0.38\pm 0.08$ & $ 0.60\pm 0.08$ \\
Q005520+123421 & 00 55 19.52 & +12 34 20.9 & $0.1783\pm0.0004$ & $20.81\pm 0.07$ & $-0.16\pm 0.10$ & $ 0.53\pm 0.11$ & $ 0.48\pm 0.09$ \\
\enddata
\tablecomments{
For each new emission line galaxy (1), the J2000 coordinates (2) and (3), the
measured redshift and error (4), the $B$ band brightness in magnitudes (5),
and the measured $U-B$ (6), $B-R$ (7), and $R-I$ (8) colors in magnitudes.
}
\label{elgs}
\end{deluxetable}


\onecolumn

\begin{figure}[hp]
  \centering
  \mbox{
\includegraphics[scale=0.9, angle=0]{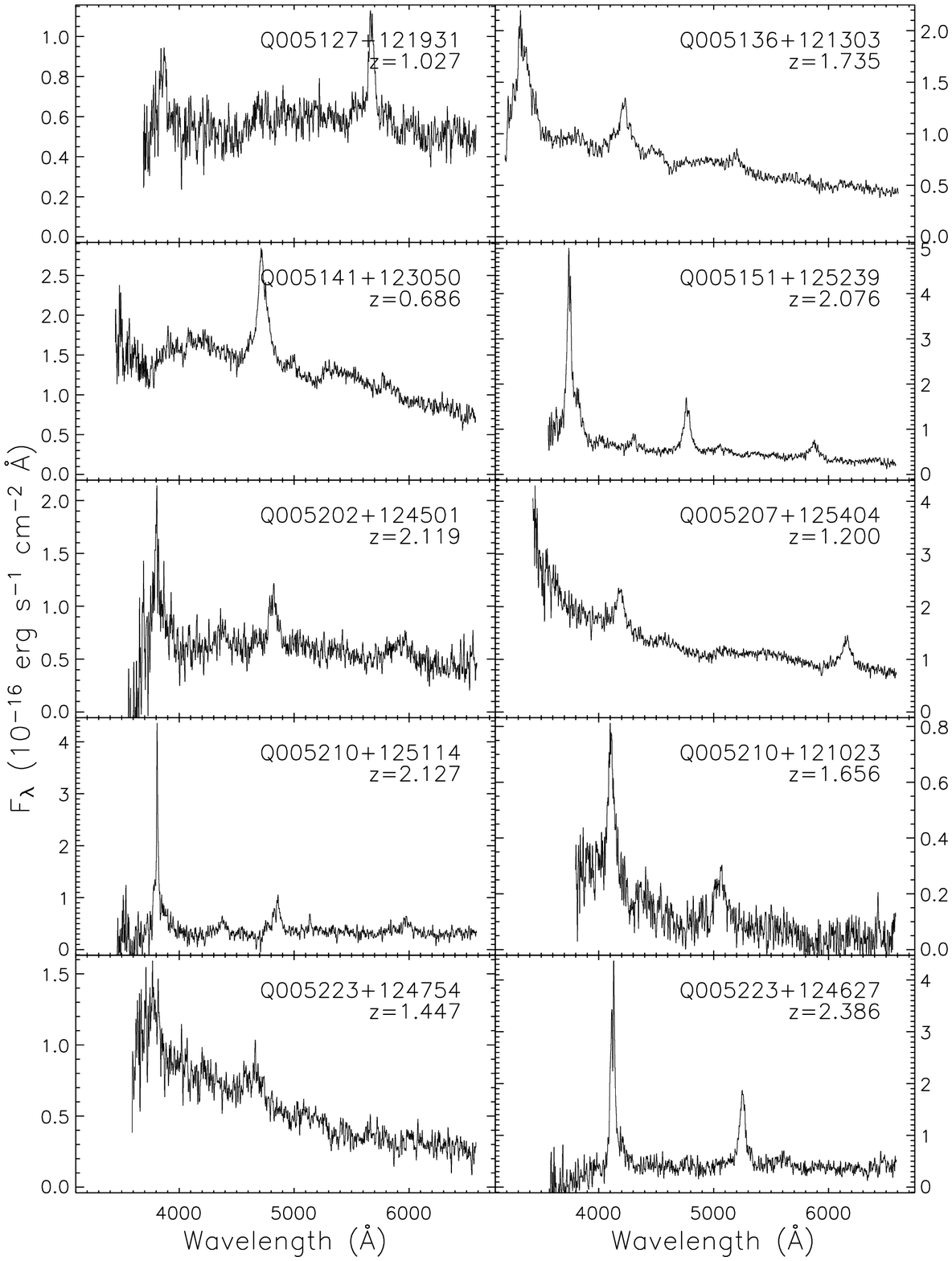}
  }
\end{figure}

\begin{figure}[hp]
  \centering
  \mbox{
\includegraphics[scale=0.9, angle=0]{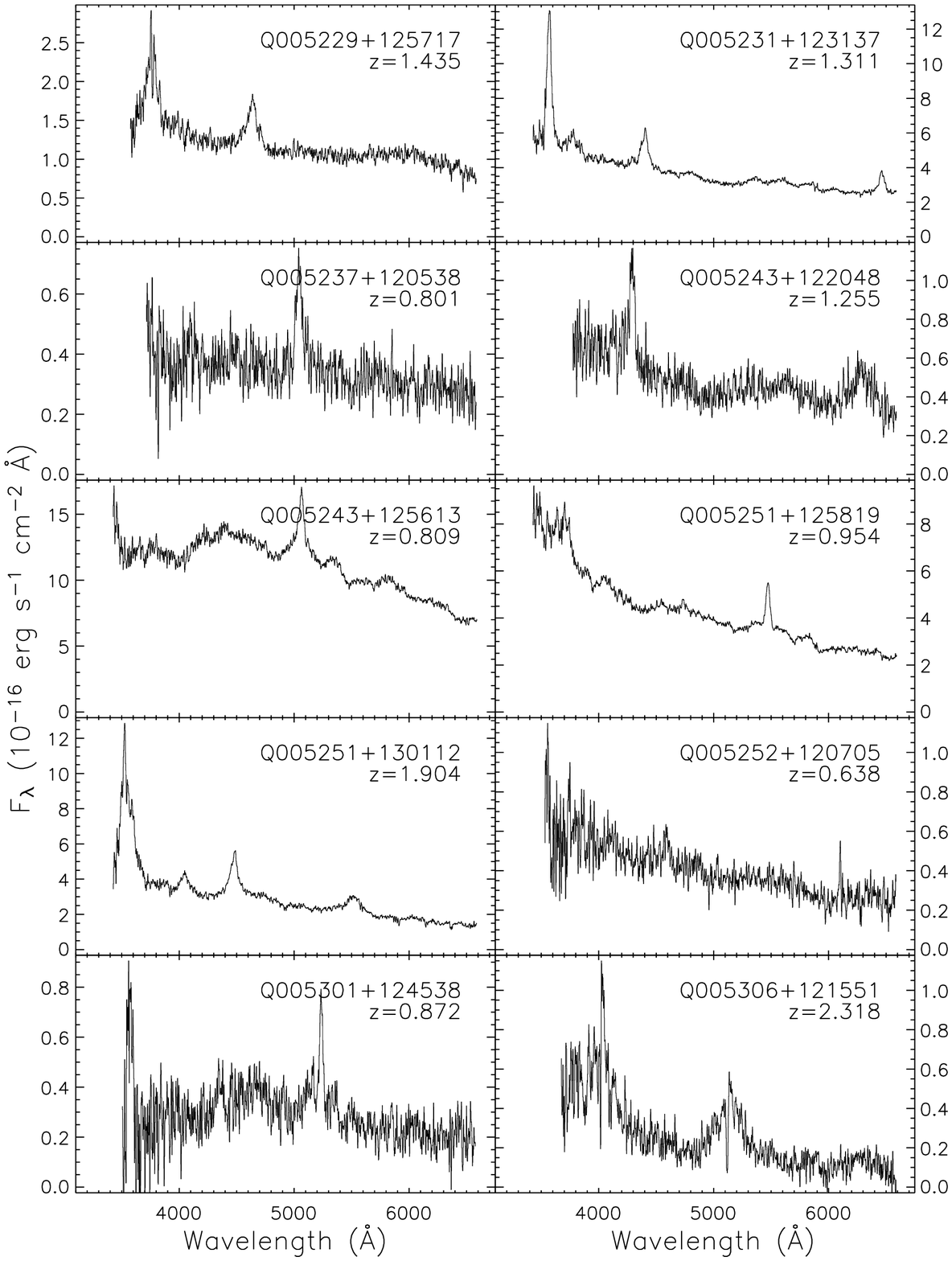}
  }
\end{figure}

\begin{figure}[hp]
  \centering
  \mbox{
\includegraphics[scale=0.9, angle=0]{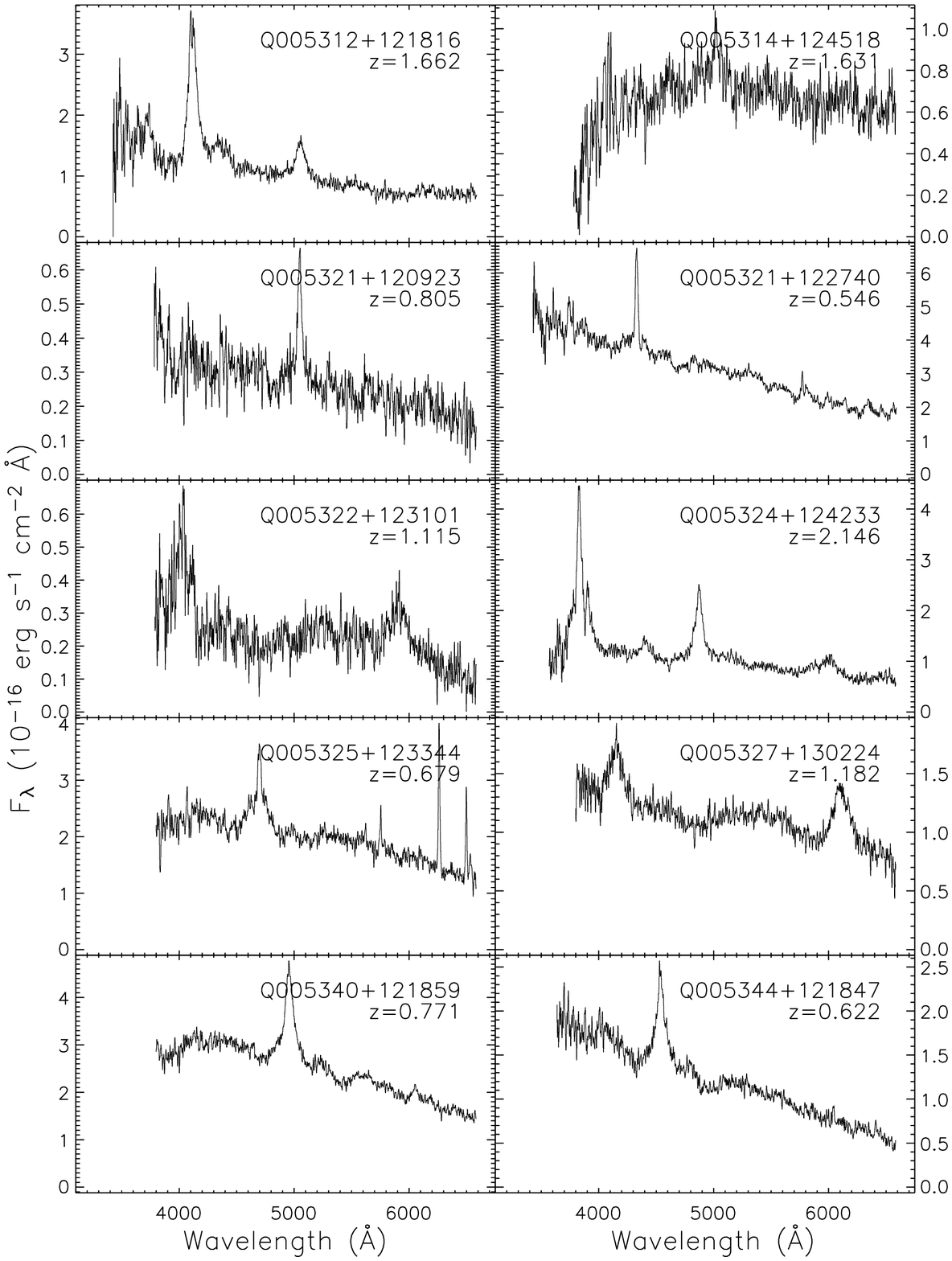}
  }
\end{figure}

\begin{figure}[hp]
  \centering
  \mbox{
\includegraphics[scale=0.9, angle=0]{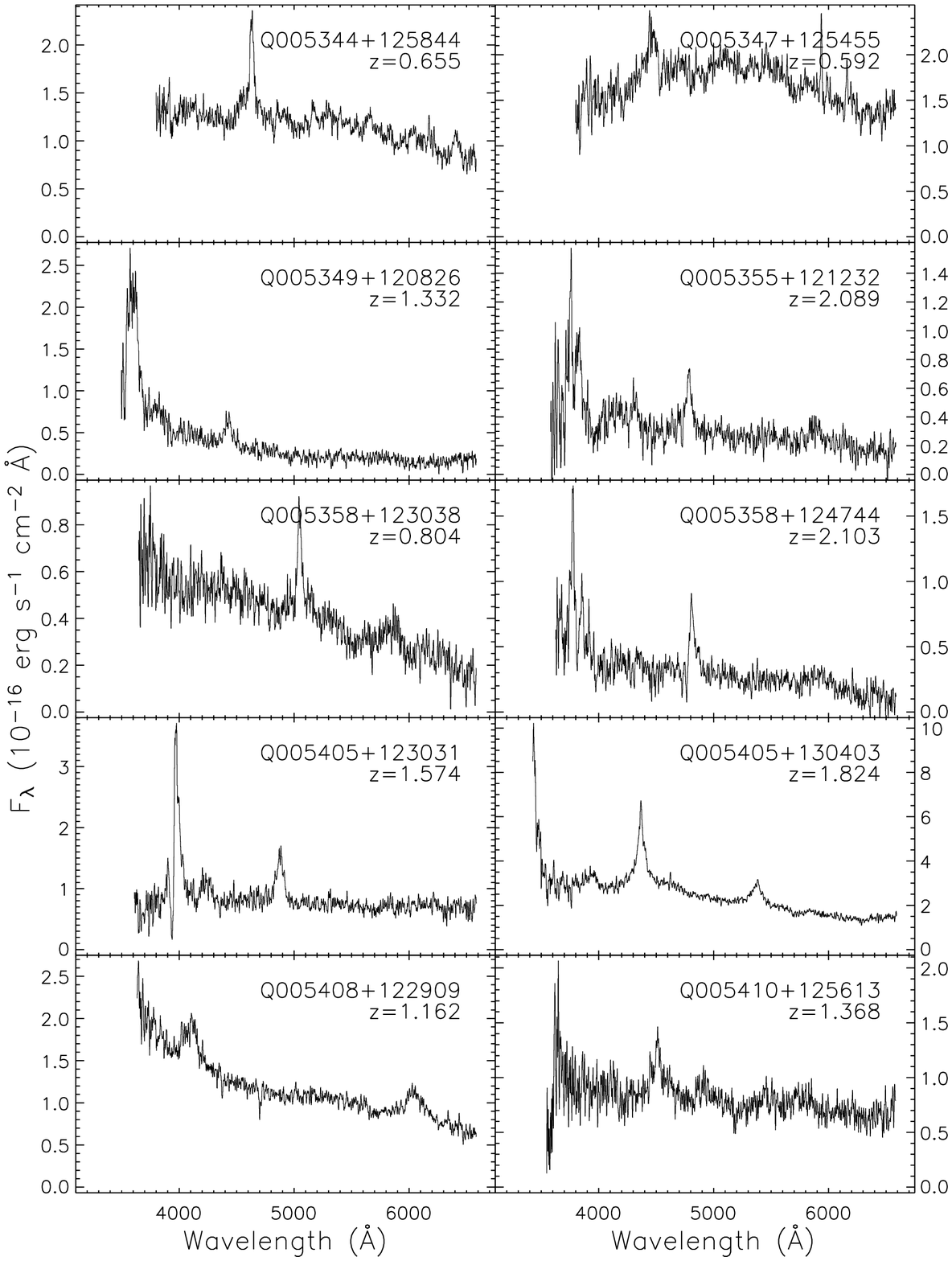}
  }
\end{figure}

\begin{figure}[hp]
  \centering
  \mbox{
\includegraphics[scale=0.9, angle=0]{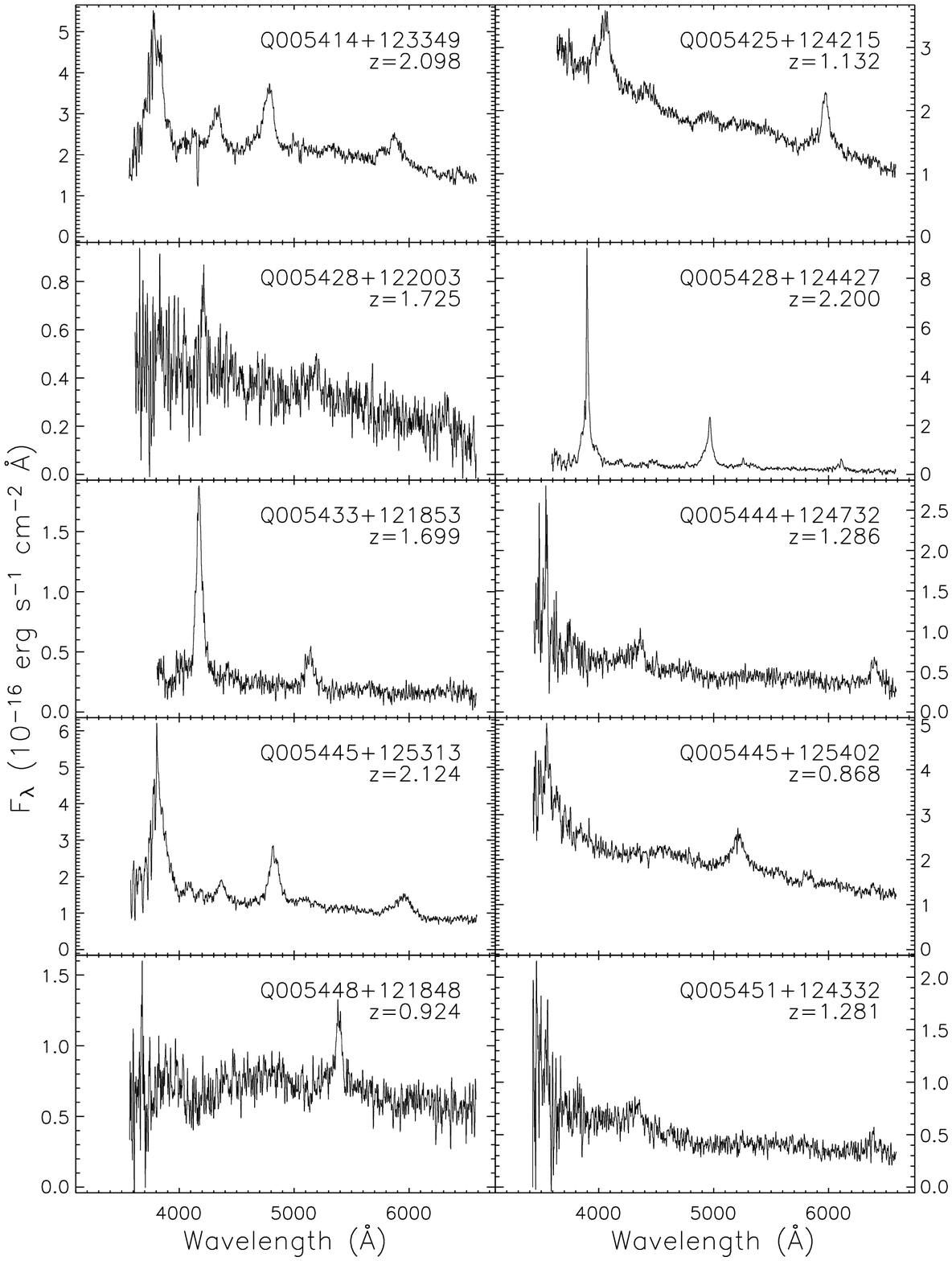}
  }
\end{figure}

\begin{figure}[hp]
  \centering
  \mbox{
\includegraphics[scale=0.9, angle=0]{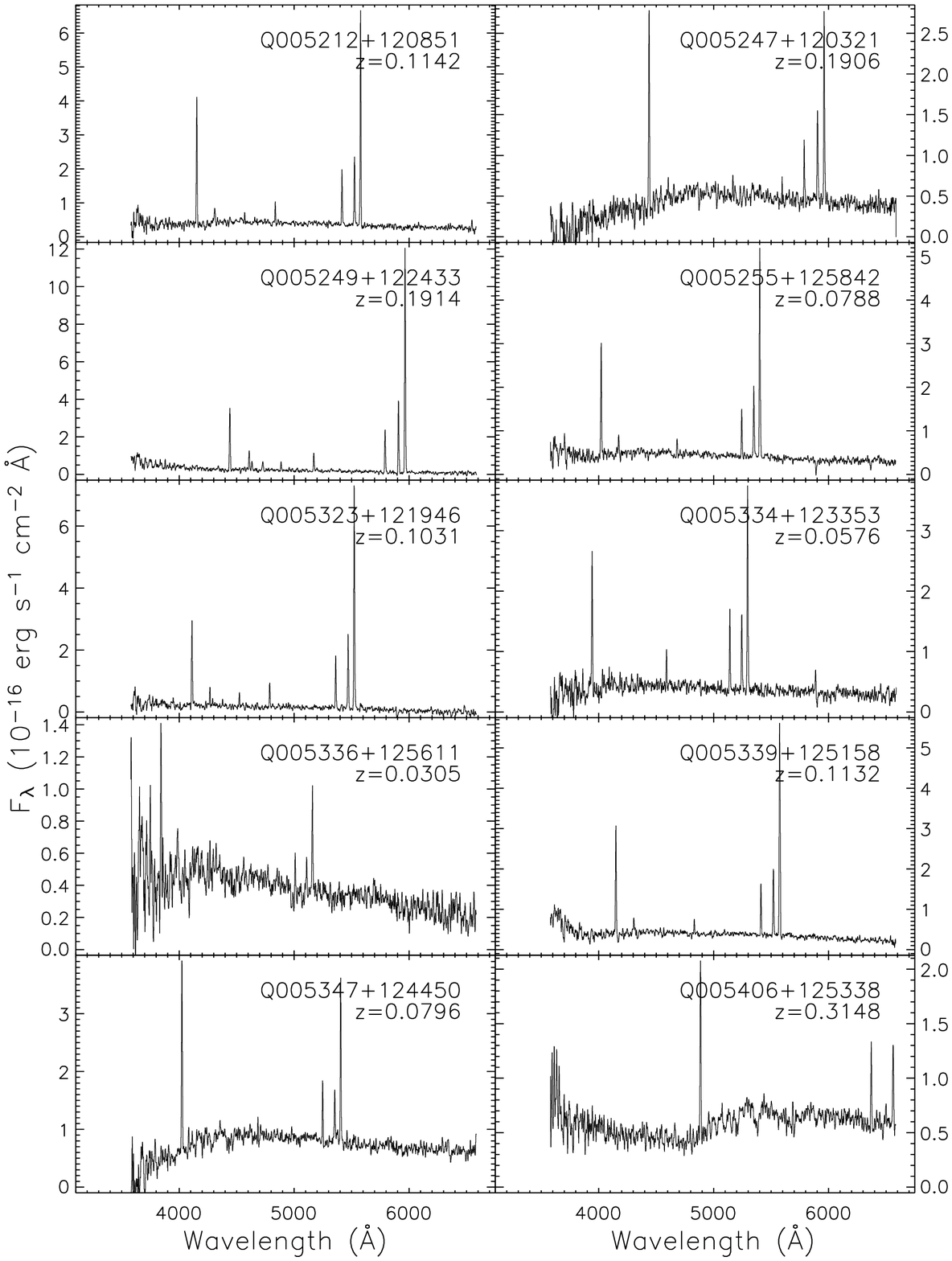}
  }
\end{figure}

\end{document}